\documentclass[prb,aps,twocolumn,showpacs,floatfix]{revtex4}

\usepackage[pdftex]{graphicx}
\usepackage{amsmath,amsfonts,amsbsy,mathrsfs,amssymb}

\begin{document}

\title{Infrared-active phonon modes in monoclinic multiferroic MnWO$_4$}

\author{T. M{\"o}ller,$^{1,2}$ P.~Becker,$^{3}$ L.~Bohat\'{y},$^{3}$ J. Hemberger,$^{1}$ and M. Gr{\"u}ninger$^{1}$}
\affiliation{$^1$ II. Physikalisches Institut, Universit\"{a}t zu K\"{o}ln, Z\"{u}lpicher Stra{\ss}e 77, D-50937 K\"{o}ln, Germany\\
$^2$ I. Physikalisches Institut, Universit\"{a}t zu K\"{o}ln, Z\"{u}lpicher Stra{\ss}e 77, D-50937 K\"{o}ln, Germany\\
$^3$ Institut f\"{u}r Kristallographie, Universit\"{a}t zu K\"{o}ln, Greinstra{\ss}e 6, D-50939 K\"{o}ln, Germany}

\date{June 24, 2014, revised: September 12, 2014}

% 63.20.kk Phonon interactions with other quasiparticles
% 63.20.-e Phonons in crystal lattices
% 78.30.-j Infrared and Raman spectra
% 77.84.-s    Dielectric, piezoelectric, ferroelectric, and antiferroelectric materials

\begin{abstract}
We report on polarized infrared reflectivity measurements of multiferroic, monoclinic MnWO$_4$ between 10\,K and 295\,K.\@
All five non-vanishing components of the dielectric tensor have
been determined in the frequency range of the phonons.
All infrared-active phonon modes (7 $A_u$ modes and 8 $B_u$ modes) are unambiguously identified.
In particular the strongest $B_u$ modes have been overlooked in previous studies,
in which the monoclinic symmetry was neglected in the analysis.
The combined analysis of reflectance data measured in different experimental geometries
($R_{ac}$ and $R_p$) is particularly helpful for a proper identification of the $B_u$ modes.
Using a generalized Drude-Lorentz model, we determine the temperature dependence of the phonon parameters,
including the orientation of the $B_u$ modes within the $ac$ plane.
The phonon parameters and their temperature dependence were discussed controversially in previous studies,
which did not include a full polarization analysis.
Our data do not confirm any of the anomalies reported above 20\,K.\@
However, in the paramagnetic phase we find a drastic reduction of the spectral weights
of the weakest $A_u$ mode and of the weakest $B_u$ mode with increasing temperature.
Below 20\,K, the parameters of the $A_u$ phonon modes for $E\, \| \, b$ show only subtle changes,
which demonstrate a finite but weak coupling between lattice dynamics and magnetism in MnWO$_4$.
A quantitative comparison of our infrared data with the quasi-static dielectric constant $\varepsilon_b$
yields a rough estimate for the oscillator strength $\Delta\varepsilon_{\rm em} \lessapprox 0.02$
of a possible electromagnon for $E\, \| \, b$.
Furthermore, we report on a Kramers-Kronig-consistent model which is able to describe
non-Lorentzian line shapes in compounds with monoclinic symmetry.
\end{abstract}

\maketitle

\section{Introduction}

The metal tungstate family $A$WO$_4$ with divalent $A$ metal ions includes a number of compounds
with interesting properties. The range of possible applications is very broad, including (phonon-) scintillation detectors,
laser wave\-guides, laser crystals, and photocatalysis.\cite{Errandonea08,Meunier99,Zhao10,Chen01,Becker07,Huang07}
The compound MnWO$_4$, also known from nature as mineral \emph{h{\"u}bnerite},
belongs to the class of multiferroics, displaying a coexistence
of antiferromagnetic and ferroelectric order parameters.\cite{Heyer06,Taniguchi06,Arkenbout06}
The Mn$^{2+}$ ions are in a high-spin 3d$^5$ configuration with spin $S$\,=\,5/2.
Magnetic frustration leads to a competition of magnetic ground states.
Upon cooling, one finds a series of magnetic phase transitions,\cite{Arkenbout06,Lautenschlaeger93}
first to an incommensurate collinear antiferromagnetic phase (AF3) at $T_{\rm N3} = 13.5$\,K,
then to an incommensurate spiral phase (AF2) at $T_{\rm N2} = 12.5$\,K,
and finally at $T_{\rm N1} \approx 6.5 - 8.0$\,K to a commensurate collinear phase (AF1).
A ferroelectric polarization and thus magnetoelectric multiferroicity is observed in the
AF2 phase\cite{Heyer06,Taniguchi06,Arkenbout06}
as well as in a further phase occurring in high magnetic fields.\cite{Nojiri11}
The spontaneous polarization is parallel to the $b$ axis.\cite{Taniguchi06,Arkenbout06}
Ferroelectricity originates from the spiral spin structure via the inverse Dzyaloshinskii-Moriya
effect.\cite{Heyer06,Taniguchi06,Arkenbout06,Katsura05,Mostovoy06,Sergienko06}
Recently, it was pointed out that competing isotropic exchange interactions are also
important for multiferroicity.\cite{Solovyev13}
The coupling between electric and magnetic effects gives rise to particularly rich physics, ranging from
the switching of the electric polarization by an external magnetic field\cite{Heyer06,Taniguchi06,Arkenbout06}
via the coupling of magnetic and electric domains\cite{Meier09}
to second-harmonic generation from an incommensurate magnetic structure\cite{Meier10}
and to a magnetoelectric memory effect.\cite{Taniguchi09,Finger10}

\begin{figure}[tb]
\includegraphics[width=\columnwidth,clip]{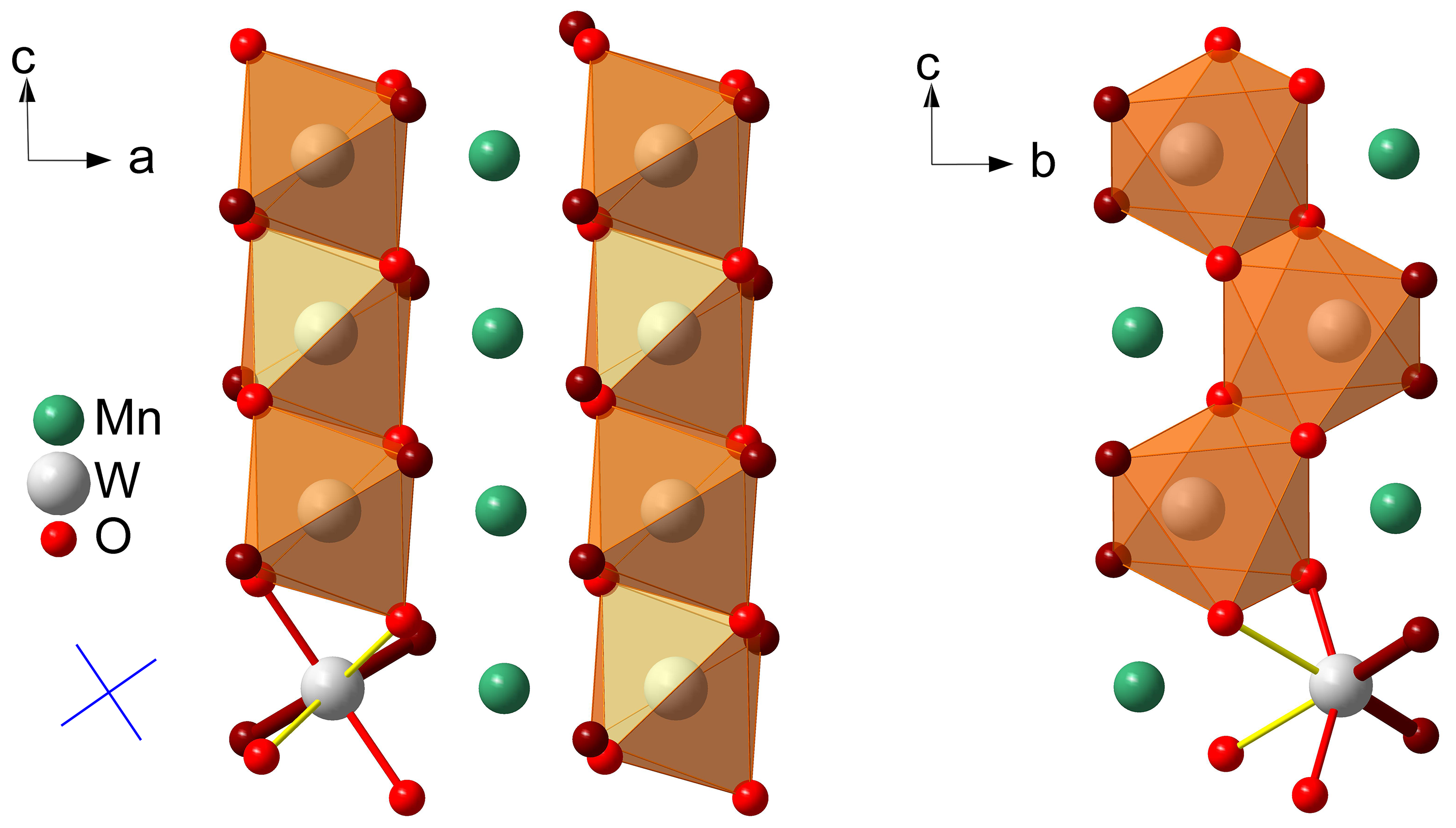}
\caption{(Color online) Sketch of the crystal structure of mono\-clinic MnWO$_4$ highlighting the chains of WO$_6$ octahedra
running along the $c$ axis. Left: $ac$ plane, right: $bc$ plane.
Dark red spheres refer to the O(2) ions with short W-O(2) bonds of only 1.79\,\AA.\@
The WO$_6$ octahedra are connected along edges via two O(1) ions (light red spheres).
Within one octahedron, the diagonals are either formed by a short W-O(2) bond (thick, dark red lines) in combination
with a long W-O(1) bond (2.13\,\AA, thin, yellow lines) or by two of the shorter W-O(1) bonds (1.91\,\AA, thin,
light red lines).
The two blue lines in the bottom left corner indicate the orientations of the two highest $B_u$ phonon modes
as derived from the infrared reflectance data. The orientations of these two modes support the interpretation
as W-O bond stretching modes. }
\label{MnWO4Struc}
\end{figure}

The magnetoelectric coupling is not restricted to static properties but is also relevant for the dynamics.
The character of magnons may change from purely magnetic to magnetoelectric, and these so-called electromagnons
can be excited by the electric field component of an electromagnetic wave,\cite{Pimenov06Nat,Katsura07,Valdes09,Kubacka14}
i.e., electromagnons contribute to the optical conductivity
and to the dielectric function $\varepsilon(\omega)$.
The spectral weight of the electromagnons has to be transferred from another dipole-active excitation.\cite{Katsura07,Yildirim08}
In the multiferroic phase of the manganites $A$MnO$_3$ (with $A$\,=\,Gd$_{1-x}$Tb$_x$ and Eu$_{1-x}$Y$_x$),
the spectral weight of the electromagnons partially stems from the phonon mode
lowest in energy.\cite{Pimenov06,Takahashi08,Schmidt09,Schleck10,Lee09,Aguilar07,Takahashi09,Shuvaev10,Takahashi12}
This behavior differs from the familiar case of proper ferroelectrics with a displacive phase transition.
There, the continuous phase transition into the polar phase is accompanied by the softening of an infrared-active phonon mode.
At the phase transition, the vanishing phonon frequency $\omega_0$ leads to a divergence of the static permittivity via
a diverging contribution to the dielectric function (or oscillator strength) $\Delta\varepsilon\propto (\omega_p/\omega_0)^2$,
where $\omega_p$ denotes the plasma frequency of the phonon. This does not require a change of the phonon's spectral weight
$\propto \omega_p^2$ in the optical conductivity.
Multiferroic MnWO$_4$ is an improper ferroelectric, in which ferroelectricity is not directly connected to a softening phonon
but rather to the onset of complex magnetic order.
In this case, the multiferroic phase transition may be accompanied by the softening of an electromagnon
as discussed for DyMnO$_3$.\cite{Shuvaev10a}
A finite spectral weight for electromagnons is expected due to the reduced symmetry in the multiferroic phase
with an order parameter of magnetoelectric origin. This spectral weight may stem from the phonons.\cite{Katsura07}
A detailed study of the lattice vibrations and of the phonon parameters thus may provide valuable information
about the ferroelectric transition and the spin-lattice coupling.

In monoclinic MnWO$_4$, several results suggest that the spin-lattice coupling is only weak.
High-resolution thermal expansion data show subtle but clear anomalies at both $T_{\rm N1}$
and $T_{\rm N3}$.\cite{Chaudhury08}
The ferroelectric polarization is of the order of 50\,$\mu$C/m$^2$,\cite{Taniguchi06,Arkenbout06}
more than an order of magnitude smaller than in, e.g., TbMnO$_3$.\cite{Kimura03}
At $T_{\rm N1}$, the static dielectric constant along the $b$ axis, $\varepsilon_b$, shows only a small
jump of roughly 0.01,\cite{Taniguchi06,Arkenbout06} which is about two orders of magnitude smaller
than the effects observed in the manganites at low frequencies.\cite{Kimura03}

The phonon modes of MnWO$_4$ have been studied by Raman scattering and optical spectroscopy.
Using polarized Raman scattering, Iliev \textit{et al.}\cite{Iliev09} found no anomalies of the phonon
parameters down to 5\,K, while Dura \textit{et al.}\cite{Dura12} reported an enhanced damping of
several phonon modes in the ferroelectric AF2 phase which was attributed to spin-phonon interactions.
In contrast, Hoang \textit{et al.}\cite{Hoang10} observed phonon anomalies at about 50\,K and between 150
and 200\,K  in their Raman data of MnWO$_4$ and suggested a new phase transition at 180\,K, far above
the known magnetic phase-transition temperatures.
As far as the infrared-active phonon modes are concerned, a consistent description is still lacking.
Choi \textit{et al.}\cite{Choi10} measured polarized reflectivity spectra for three different polarization
directions of the electric field $E$ ($E\, \| \, a$, $b$, and $c$), which is not sufficient
for a proper identification of the eigenmodes in this monoclinic compound where $a$ and $c$
are not mutually perpendicular.
To the best of our knowledge, an analysis of the full dielectric tensor in the frequency
range of the phonons has not been reported for any mono\-clinic tungstate $A$WO$_4$ with
divalent $A$ metal ions thus far.
Choi \textit{et al.}\cite{Choi10} found no anomalies of the phonons as a function of temperature and
reported only representative data on the temperature dependence for two high-energy features
above 80\,meV.\@

Remarkably, stronger anomalies were reported both in Raman and infrared data for polycrystalline samples
of Mn$_{1-x}A_x$WO$_4$ doped with a few percent of $A$\,=\,Fe, Co, or Ni.\cite{Maczka11,Ptak12}
Most of these anomalies were observed between 20\,K and 200\,K,
i.e., far above the magnetic phase-transition temperatures of undoped MnWO$_4$.
In the case of $A$\,=\,Co, high-resolution synchrotron X-ray diffraction data\cite{Urcelay12} for
$x$\,=\,0.05 and 0.20 show only small anomalies at the magnetic phase transitions.
Moreover, the reported eigenfrequencies of the slightly doped samples\cite{Maczka11,Ptak12}
strongly deviate from those reported for pure MnWO$_4$.\cite{Choi10} None of these studies
takes the monoclinic structure fully into account. For monoclinic symmetry, optical phonon modes
in general show a mixture of transverse (TO) and longitudinal (LO) character,
and this mixture depends on the direction of the wavevector $k$.
Due to the LO-TO splitting, the eigenfrequency of a given phonon mode also
depends on the direction of $k$, thus the apparent peak frequency
varies with the geometry of the experiment.\cite{Koch74,Kuzmenko01}
This explains the difficulties in determining the correct eigenfrequencies, in particular for polycrystalline samples.

Further infrared studies were performed on nanocrystalline MnWO$_4$ with different morphologies of the
nanoparticles.\cite{Maczka11b,Maczka12,Tong10}  In contrast to Raman modes, the infrared-active phonons show a
pronounced dependence on particle size and morphology.\cite{Maczka11b,Maczka12} As stated above, this is
not surprising since the TO-LO mixture and thus the mode frequency depend on the direction of $k$.\cite{Koch74}
It has been concluded\cite{Maczka11b} that a detailed understanding of the bulk modes is a prerequisite
for a correct description of the phonon modes of nanocrystals.

Here, we report on a full polarization analysis of single-crystalline MnWO$_4$,
which allows us to identify unambiguously all expected phonon modes, both for $A_u$ and $B_u$ symmetry.
Using a generalized Drude-Lorentz model, we determine the temperature dependence of
all phonon parameters, including the orientation of the $B_u$ modes within the $ac$ plane.
A comparison to previous studies\cite{Choi10,Maczka11,Ptak12} shows that in particular the
strongest modes have been overlooked thus far. This surprising result can be explained easily.
Weaker modes show a small LO-TO splitting and thus give rise to rather narrow but clear features.
Strong modes with a very large LO-TO splitting yield broad features, and the
orientational dispersion of
the dielectric tensor gives rise to unusual line shapes of these broad peaks.
Moreover, the TO-LO mixture depends on the direction of the wavevector
$k$; thus the eigenfrequency of modes with a large LO-TO splitting may change strongly as a function of $k$.

The paper is organized as follows.
Experimental details are given in Sec.\ II, followed in Sec.\ III by a factor-group analysis.
Section IV describes the dielectric tensor in monoclinic symmetry as well as the models used
to analyze the infrared data, i.e., a generalized Drude-Lorentz model (Sec.\ IV A),
an asymmetric, Kramers-Kronig-consistent oscillator model (Sec.\ IV B), and a
Kramers-Kronig-constrained variational approach (Sec.\ IV D).
The reflectivities $R_b$, $R_{ac}$, and $R_p$ measured in different experimental geometries
are introduced in Sec.\ IV C.\@
Section V describes our results for the phonon modes.
In Sec.\ V A and V B we address the $A_u$ and $B_u$ modes, respectively,
followed by a detailed discussion of the line shape of the highest $B_u$ mode in Sec.\ V C.\@
Finally, the temperature dependence and the transfer of spectral weight from the phonons
to either lower or higher frequencies are discussed in Sec.\ V D.\@
Conclusions are given in Sec.\ VI.

\section{Experiment}
\label{SecExp}
Single crystals of MnWO$_4$ were grown from the melt using the top-seeding technique.
The Mn ions can be kept in the divalent state during growth by using a high growth
temperature and avoiding melt solvents.\cite{Becker07}
We obtained ruby-red transparent crystals with dimensions up to $5\, \times \, 5 \, \times \, 25\,$mm$^3$.
The crystal structure\cite{Weitzel76,Macavei93} of MnWO$_4$ is monoclinic with space group $P2/c$,
the monoclinic angle amounts to $\beta =91.08^\circ$.
Edge-sharing distorted [MnO$_6$] octahedra and edge-sharing distorted [WO$_6$] octahedra
form alternating zig-zag chains running along the $c$ axis, see Fig.\ \ref{MnWO4Struc}.
We used natural growth faces and Laue diffraction for the crystallographic sample orientation.
After orientation, the samples were lapped and polished.

\begin{figure}[b]
\includegraphics[width=0.85\columnwidth,clip]{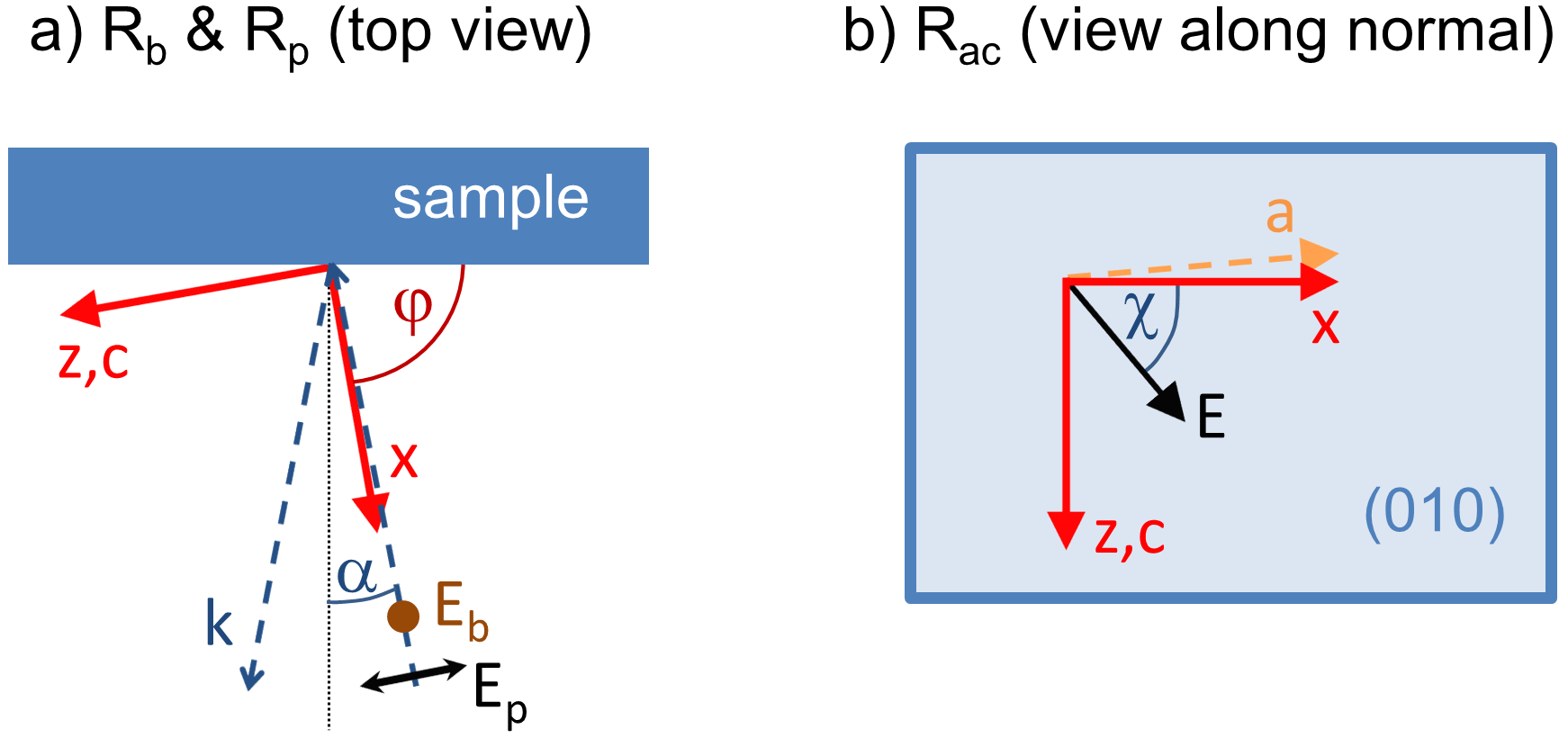}
\caption{(Color online) Sketches of the experimental geometries for measuring
a) $R_b(\omega)$ and $R_p(\omega,\alpha,\varphi)$ with (010) as plane of incidence,
and b) $R_{ac}(\omega,\chi)$ on a (010) surface.
In both cases, the view is along the $b$ axis with $b\, \| \, y$.
Red: Cartesian coordinates $x$ and $z\, \| \, c$.
a) Dotted: surface normal. Dashed: wave vectors of incident and reflected light.
$\alpha$\,=\,11$^\circ$ denotes the angle of incidence, and $\varphi$\,=\,80$^\circ$
is the angle between the $x$ axis and the sample surface.
$R_b$ was measured with $s$-polarized light, $R_p$ with $p$-polarized light.
b) For the analysis of $R_{ac}(\omega,\chi)$ we assume normal incidence.
$\chi$ denotes the angle by which the polarization direction of the electric field $E$
has to be rotated to coincide with the $x$ axis.
The angle $\beta$\,=\,91.08$^\circ$ between $a$ and $c$ axes is exaggerated for clarity.
}
\label{geometry}
\end{figure}

Using a \textsc{Bruker} IFS 66v/S Fourier-transform spectro\-meter, we performed reflectivity measurements
at nine different temperatures ranging from
10\,K to 295\,K in the frequency range of 50 - 7000 cm$^{-1}$.
The sample was mounted on the cold finger of a continuous-flow He cryostat.
The angle of incidence $\alpha$ was about $11^\circ$, i.e., near-normal incidence.
The incident light was linearly polarized, while the polarization state of the reflected
light was not analyzed.
The temperature of the sample was measured with a thermometer glued on the sample,
avoiding direct thermal contact between thermometer and sample holder.
We report data down to 20\,K for the paramagnetic phase, for 13\,K in the AF3 phase,
and for 10\,K in the multiferroic AF2 phase. The phase transition to the commensurate collinear
AF1 phase at 6.5 - 8.0\,K could not be reached.
Reference measurements were obtained using \emph{in-situ} Au evaporation.

In monoclinic MnWO$_4$ the $b$ axis is perpendicular to the $ac$ plane.
We use a Cartesian coordinate system with $y \, \| \, b$, $z \, \| \, c$,
and $x$ lying in the $ac$ plane with $x \! \perp \! z$, see Fig.\ \ref{geometry}b).
The reflectivity $R_b(\omega)$ was measured on a surface containing the $b$ axis with polarization
of the electric field $E \, \| \, b$.
To avoid any contribution from the $ac$ plane, we chose the (010) plane as plane of incidence,
i.e., $R_b(\omega)$ was measured with $s$-polarized light [$E \! \perp \! $ to (010)] with an angle
of incidence of $\alpha$\,=\,11$^\circ$, see Fig.\ \ref{geometry}a).
The sample surface deviates from a (100) surface by 10$^\circ$,
i.e., $\varphi$\,=\,80$^\circ$ denotes the angle by which
the $x$ axis has to be rotated around the $y$ axis to coincide with the sample surface.
In the same geometry, we measured the reflectivity $R_p (\omega,\alpha,\varphi)$ for $p$-polarized light,
i.e., with $E \! \perp \! b$.

\begin{figure}[t]
\includegraphics[width=0.9\columnwidth,clip]{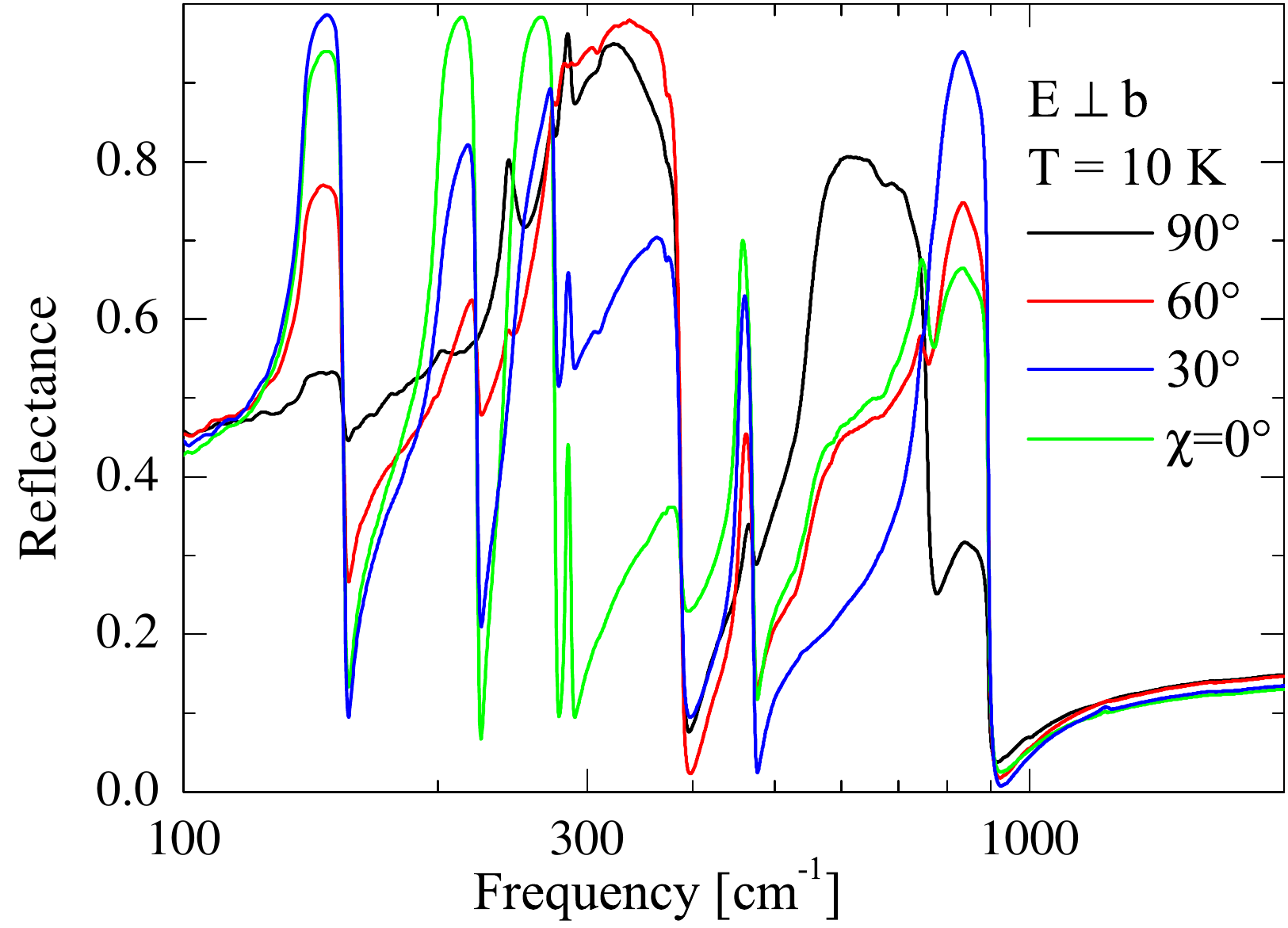}
\caption{(Color online) Reflectance $R_{ac}(\omega,\chi)$ of MnWO$_4$ for $E \! \perp \! b$
measured on a (010) surface for four different polarization angles $\chi$ at $T$\,=\,10\,K,
cf.\ Fig.\ \ref{geometry}b).
Note the logarithmic frequency scale.
}
\label{MnWO4RefAllAngles}
\end{figure}

\begin{figure}[t]
\includegraphics[width=0.9\columnwidth,clip]{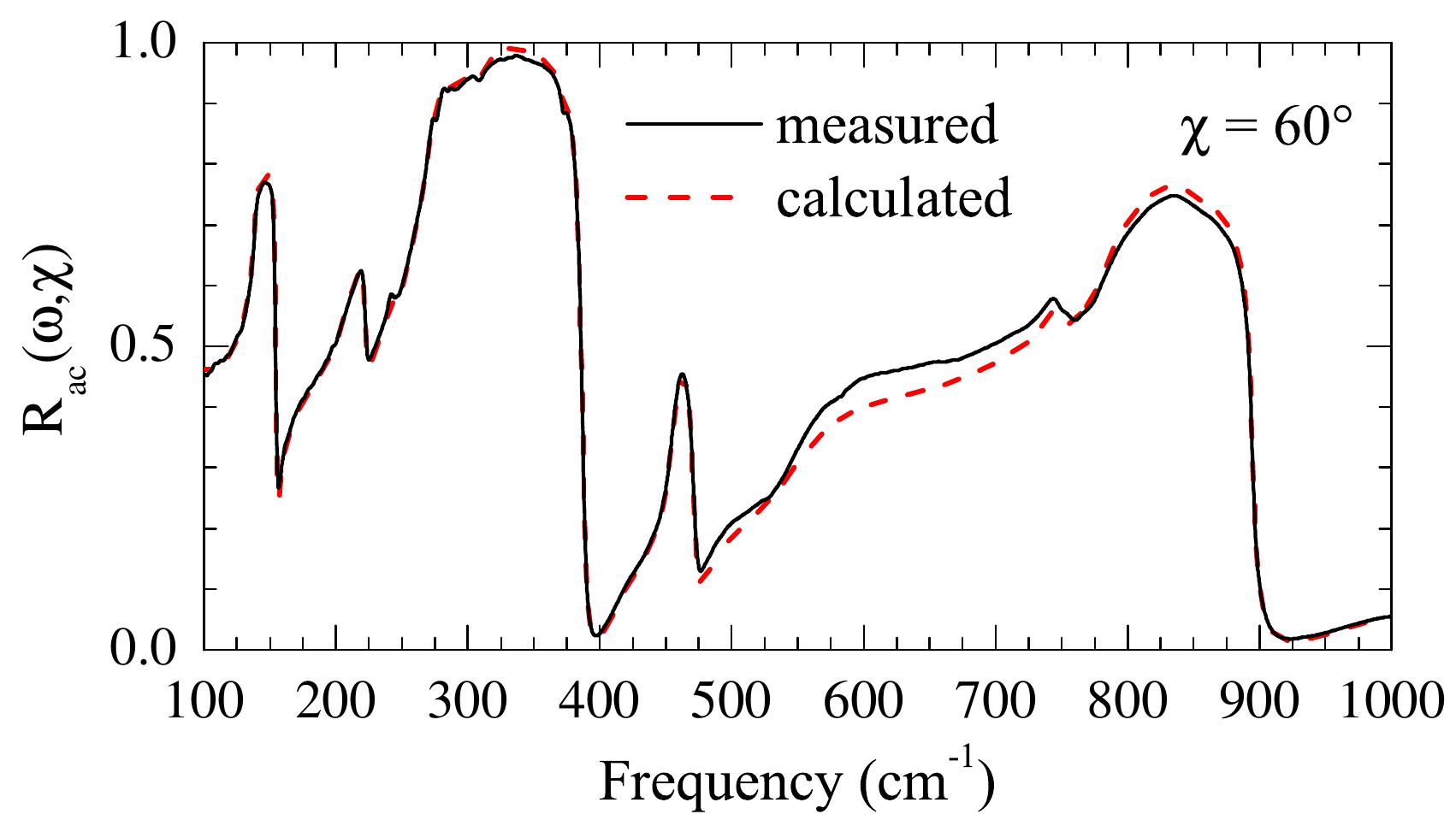}
\caption{(Color online) Solid: measured reflectivity $R_{ac}(\omega,\chi)$ for $\chi$\,=\,60$^\circ$ at $T$\,=\,10\,K.\@
Red dashed: calculated reflectivity
$R_{ac}^{\rm calc}(60^\circ) = -\frac{1}{2} R_{ac}(0^\circ) + R_{ac}(30^\circ) +\frac{1}{2} R_{ac}(90^\circ)$.
The maximum deviation between measured and calculated data amounts to 5\,\% at about 600\,cm$^{-1}$.
}
\label{MnWO4Ref60deg}
\end{figure}

For a full polarization analysis of this monoclinic compound, we measured the reflectivity $R_{ac}(\omega,\chi)$
on a (010) surface, where $\chi$ denotes the angle between the incident electric field $E$ and the $x$ axis,
see Fig.\ \ref{geometry}b). For simplicity,
we assume normal incidence for the definition of $\chi$ and for the analysis of $R_{ac}(\omega,\chi)$.
The polarization direction $\chi$ was varied by rotating not the sample but the polarizer using a stepper motor.
This bears the advantages that the angular precision is higher and that the polarization angle can be changed
while the sample is kept at low temperature. The disadvantage is that the incident electric field is not exactly
parallel to the $ac$ plane (with the exception of $s$-polarized light) due to the finite angle of incidence.
We measured $R_{ac}(\omega,\chi)$ for $\chi = 0^\circ$, $30^\circ$, $60^\circ$, and $90^\circ$,
see Fig.\ \ref{MnWO4RefAllAngles}. Any three of these data sets can be used to calculate $R_{ac}(\omega,\chi)$
for any value of $\chi$.\cite{Kuzmenko96} The comparison of the measured reflectivity for $\chi$\,=\,60$^\circ$
with the calculated one in Fig.\ \ref{MnWO4Ref60deg} demonstrates the consistency of our data.

Additionally, we measured the real part of the quasi-static dielectric constant along the $b$ axis,
Re$\{\varepsilon_b\}$, between 5\,K and 50\,K
at 96.8\,kHz and 45\,MHz.
At 96.8\,kHz we employed a frequency-response analyzer {(\sc Novocontrol)}
and a small single crystal of MnWO$_4$ with dimensions of about $2\times 0.5 \times 2$\,mm$^3$
which was prepared as a plate-type capacitor using silver-paint electrodes on the $\{010\}$ surfaces.
At 45\,MHz we used a micro-strip setup and a vector network analyzer {(\sc Rohde \& Schwarz)}.
The quasi-static data show a high relative accuracy. Here, we use the results from the infrared data
to fix the absolute value of Re$\{\varepsilon_b\}$.

\section{Factor-group analysis}

\begin{table}[t]
  \centering
  \begin{tabular}{cccc}
    \hline
         & Wyckhoff &   site   & irreducible     \\
    atom & notation & symmetry & representations \\
    \hline
    Mn   & 2(f)     & $C_2$    & $A_g$ + $A_u$ +  $2B_g$ +  $2B_u$\\
    W    & 2(e)     & $C_2$    & $A_g$ + $A_u$ +  $2B_g$ +  $2B_u$\\
    O(1) & 4(g)     & $C_1$    & $3A_g + 3A_u + 3B_g + 3B_u$\\
    O(2) & 4(g)     & $C_1$    & $3A_g + 3A_u + 3B_g + 3B_u$\\
    \hline
    \hline
  \end{tabular}
  \caption{Atomic site symmetries\cite{Becker07} and irreducible representations
  for the atoms in monoclinic MnWO$_4$ with space group $P2/c$.}
  \label{MnWO4SiteSymmetries}
\end{table}

The number of phonon modes can be derived from a factor-group analysis.\cite{Rousseau}
Monoclinic MnWO$_4$ with space group $P2/c$ has $Z$\,=\,2 formula units per unit cell.
For $T$\,=\,295\,K, the site symmetries as well as the irreducible representations
of each atomic site are given in Tab.\ \ref{MnWO4SiteSymmetries}.\@
In the presence of a center of inversion, Raman activity and infrared activity of
normal modes are mutually exclusive.
In total the irreducible representations contain 36 modes,
\begin{equation}
    \Gamma_{\rm total} = 8 \, A_g \, + \, 8 \, A_u \, + \, 10 \, B_g \, + \, 10
    B_u.
\end{equation}
Subtracting the acoustic modes $(A_u + 2 \, B_u)$ and the Raman modes
$(8 \, A_g \, + \, 10 \, B_g)$, we find 15 infrared-active phonon modes
\begin{equation}
\label{AuBu}
    \Gamma_{\rm IR} = 7 \, A_u \, + \, 8 \, B_u.
\end{equation}
The factor-group analysis thus predicts seven $A_u$ modes for polarization of the
electric field $E \, \| \, b$ and eight $B_u$ modes for polarization within the $ac$ plane, $E \! \perp \! b$.
The magnetic phase transition at $T_{\rm N3}$ is not connected with a structural phase
transition,\cite{Lautenschlaeger93} thus our analysis is valid down to $T_{\rm N2}$\,=\,12.5\,K.\@
Below $T_{\rm N2}$, the appearance of ferroelectricity reflects the loss of a mirror plane
and a concurrent change of the selection rules. Thus, the distinction between Raman-active and
infrared-active modes is not strictly valid anymore.
However, the ferroelectric polarization $P_b$ is only small\cite{Heyer06,Taniguchi06,Arkenbout06}
in MnWO$_4$, about 50\,$\mu$C/m$^2$.
Hence the ionic displacements $\delta u \propto P_b$ are expected to be small,
and they have escaped detection in structural studies so far.\cite{Lautenschlaeger93}
Accordingly, we expect that the Raman-active modes acquire only a tiny oscillator strength $\Delta \varepsilon \propto P_b^2$
in the dielectric function\cite{Aguilar06TbMn2O5} below $T_{\rm N2}$, possibly below the detection limit.

\section{Dielectric response of a monoclinic compound}
\label{SecDielec}

In monoclinic MnWO$_4$, the tensor of the dielectric function $\hat{\varepsilon} (\omega)$ has the following form:
\begin{equation}\label{EpsTensor}
    \hat{\varepsilon} (\omega) = \left(
      \begin{array}{ccc}
        \varepsilon_{xx}(\omega) & 0                        & \varepsilon_{xz}(\omega)\\
        0                        & \varepsilon_{yy}(\omega) & 0\\
        \varepsilon_{xz}(\omega) & 0                        & \varepsilon_{zz}(\omega)\\
       \end{array}
     \right).
\end{equation}
Here, we used the Cartesian coordinate system defined in Sec.\ \ref{SecExp} with $y \, \| \, b$, $z \, \| \, c$,
and $x$ lying in the $ac$ plane (see Fig.\ \ref{MnWO4RefAllAngles}).
In the absence of an external magnetic field and neglecting a possible magnetization,
the off-diagonal matrix elements $\varepsilon_{xz}$ and $\varepsilon_{zx}$ are equal.
We decompose the three-dimensional tensor $\hat{\varepsilon} (\omega)$
into a scalar $\varepsilon_b(\omega) = \varepsilon_{yy}(\omega)$ and a two-dimensional tensor
\begin{equation}\label{EpsACTensor}
    \hat{\varepsilon}_{ac} (\omega) = \left(
      \begin{array}{cc}
        \varepsilon_{xx}(\omega) & \varepsilon_{xz}(\omega)\\
        \varepsilon_{xz}(\omega) & \varepsilon_{zz}(\omega)\\
       \end{array}
     \right).
\end{equation}
The scalar $\varepsilon_b(\omega)$ contains information on the $A_u$ modes and can be studied
by measuring the reflectivity $R_b(\omega)$ with $E \, \| \, b$ and a subsequent analysis using
a Drude-Lorentz model (see Sec.\ \ref{SecDruLo}).
For the analysis of $\hat{\varepsilon}_{ac} (\omega)$ and the modes with $B_u$ symmetry,
we employed three different methods:
a generalized Drude-Lorentz model\cite{Kuzmenko01,Kuzmenko96} (see Sec.\ \ref{SecDruLo}),
an oscillator model which takes an asymmetric (non-Lorentzian) line shape into account (see Sec.\ \ref{SecAsymm}),
and a Kramers-Kronig-constrained variational analysis\cite{Kuzmenko05}
(KKvar, see Sec.\ \ref{SecKK}).
The two-dimensional tensor $\hat{\varepsilon}_{ac}(\omega)$ can be determined
by measuring the reflectivity $R_{ac}(\omega,\chi)$ with $E \! \perp \! b$ for three different polarization
directions $\chi$ (see Sec.\ \ref{SecExp} and Fig.\ \ref{geometry}).
For the analysis of $R_{ac}(\omega,\chi)$, we assume normal incidence and that
only transverse modes are excited by the incident wave.
For $E\! \perp \! b$, the excitation of purely transverse modes actually requires that $k \, \| \, b$,
i.e., strictly normal incidence on the (010) plane. \textit{A posteriori}, this assumption
of purely transverse excitations is validated by our analysis with the exception of the $B_u$ mode
highest in energy, which is nearly degenerate with a longitudinal mode and shows a non-Lorentzian line shape
(see Sec.\ \ref{SecLineShape}).
We use the terms \textit{transverse mode} and \textit{longitudinal mode} for, e.g., the
discussion of a strong Reststrahlen band (with reflectance close to 1) in $R_{ac}(\omega,\chi)$,
even though a strict distinction between TO and LO modes is generally not valid for monoclinic symmetry.

\subsection{Generalized Drude-Lorentz model}
\label{SecDruLo}

To determine the scalar $\varepsilon_b (\omega)$ from the measured reflectivity data in the frequency range of the
phonon modes, $\varepsilon_b (\omega)$ can be described by a sum of oscillators. We employ a Drude-Lorentz model
\begin{equation}
\label{EPSb}
\varepsilon_b (\omega) = \varepsilon_b^{\infty} +
\sum_{i,A_u} \frac{\omega_{p,i}^2}{\omega_{0,i}^2 - \omega^2 - i \gamma_i \omega},
\end{equation}
where $\varepsilon_b^{\infty}$ denotes the high-frequency dielectric constant,
$\omega_{0,i}$ the transverse eigenfrequency, $\omega_{p,i}$ the plasma frequency, and
$\gamma_i$ the damping of the $i$-th oscillator, where $i$ is running over all $A_u$ modes.
The oscillator strength is given by $\Delta\varepsilon_i \! = \! (\omega_{p,i}/\omega_{0,i})^2$.

The parametrization of $\hat{\varepsilon}_{ac} (\omega)$ is somewhat more difficult because the axes $a$ and $c$
are not mutually perpendicular.
For any frequency, one can find a set of orthogonal axes within the $ac$ plane
that yields a diagonal form of the real part of $\hat{\varepsilon}_{ac}$
and a second set of axes for which the imaginary part is diagonal.
In the case of an orthorhombic crystal, the two sets of axes coincide and are fixed with respect to the crystallographic axes.
In the case of monoclinic MnWO$_4$,
symmetry fixes only the $y$ axis of the tensor $\hat{\varepsilon}$
with respect to the crystallographic axes, the two other axes may rotate within the $ac$ plane.
The rotation angles $\phi_{\rm Re}(\omega)$ and $\phi_{\rm Im}(\omega)$ of the principal axes of
Re$\left\{ \hat{\varepsilon}_{ac}(\omega) \right\}$ and Im$\left\{ \hat{\varepsilon}_{ac}(\omega) \right\}$
may differ from each other and depend on the frequency $\omega$.
This orientational dispersion of the principal axes of the dielectric tensor usually
gives rise to so-called axial dispersion, i.e., an orientational dispersion of the optical axes.
For a given frequency $\omega_0$,
$\phi_{\rm Re}(\omega_0) \neq \phi_{\rm Im}(\omega_0)$ implies that the complex tensor $\hat{\varepsilon}_{ac} (\omega_0)$
cannot be diagonalized by a rotation. In terms of an oscillator model,
the orientational dispersion of $\hat{\varepsilon}_{ac}(\omega)$ can be described
by assigning a particular orientation to each oscillator.
This generalized Drude-Lorentz (gDL) model\cite{Kuzmenko96,Kuzmenko01} reads
\begin{equation}\label{EPSacMonoclinic}
    \begin{split}
        \hat{\varepsilon}_{ac}^{\rm \, gDL} (\omega) &= \hat{\varepsilon}_{ac}^{\infty}
            + \sum_{i,B_u} \frac{\omega_{p,i}^2}{\omega_{0,i}^2 - \omega^2
            - i \gamma_i \omega} \\
         & \qquad \qquad \times  S(\theta_i) \left(
            \begin{array}{cc}
              1  & 0\\
              0 & 0 \\
            \end{array}
         \right) S^{-1}(\theta_i) \,\, ,
  \end{split}
\end{equation}
where $\hat{\varepsilon}_{ac}^{\infty}$ is a symmetric real two-dimensional tensor denoting the
high-frequency contribution, $\theta_i$ is the angle by which the dipole moment of mode $i$
has to be rotated to coincide with the $x$ axis, and $S(\theta_i)$ is the rotation matrix
\begin{equation}
            S(\theta_i)= \left( \begin{array}{cc}
              \cos \theta_i & -\sin \theta_i\\
              \sin \theta_i & \cos \theta_i \\
            \end{array}\right) \,\, .
\label{rotmat}
\end{equation}
Note that $\theta_i$ and $\phi_{\rm Im}(\omega_{0,i})$ or $\phi_{\rm Re}(\omega_{0,i})$
do not necessarily coincide. In the case of a weak mode,
$\phi_{\rm Im}(\omega_{0,i})$ may be dominated by a stronger mode which is close in energy, and
thus $\theta_i$ and $\phi_{\rm Im}(\omega_{0,i})$ may differ significantly from each other.\cite{Goessling05}

\subsection{Asymmetric oscillator model}
\label{SecAsymm}

On the whole, the generalized Drude-Lorentz model yields a satisfactory description of the reflectance of MnWO$_4$.
However, the highest $B_u$ phonon mode shows an unusual line shape in $R_{ac}(\omega,\chi)$ (see Sec.\ \ref{SecLineShape}).
In the case of a \emph{scalar} dielectric function, an asymmetric line shape [or, more precisely, a non-Lorentzian
line shape of $\varepsilon(\omega$)] can be described using a factorized four-parameter
model\cite{Chaves,Berreman,Maczka11} which employs two different values $\gamma_{T,i}$ and $\gamma_{L,i}$
for the damping of the $i$-th oscillator at the transverse and longitudinal eigenfrequencies. This mimics
an approximately quadratic frequency dependence of the damping $\gamma = \gamma(\omega)$.
However, the condition $\gamma_{L,i} > \gamma_{T,i}$ has to be satisfied,\cite{Lowndes} thus the factorized model
is capable of describing an asymmetric mode which is steep at low frequencies and washed out at the high-frequency side.
Our data show the opposite behavior (see below).
Moreover, the factorized model describes a \emph{scalar} dielectric function and cannot be written as a
sum of individual oscillators, thus the generalization of the factorized model for monoclinic symmetry
with
orientational dispersion of $\hat{\varepsilon}_{ac}(\omega)$ is not straightforward.

In the case of the tensor $\hat{\varepsilon}_{ac}(\omega)$, we choose a different approach for the description
of a non-Lorentzian line shape, starting from a sum of oscillators as described in Eq.\ \ref{EPSacMonoclinic}.
A thorough discussion of the frequency dependence of the damping $\gamma(\omega)$ requires to treat $\gamma(\omega)$ as a
response function to keep $\hat{\varepsilon}_{ac}(\omega)$ Kramers-Kronig consistent, i.e., to obey causality.
To this end, we consider the coupling between two oscillators as discussed by Barker and Hopfield\cite{Barker64}
(a simplified version valid in a narrow frequency range has been proposed by Huml\'{i}\v{c}ek {\it et al.}\cite{Humlicek00}).
More precisely, we study the coupling of one infrared-active mode (IR) and one IR-silent mode (s)
with vanishing effective charge to describe the unusual line shape of the highest $B_u$ mode.
A possible candidate for the IR-silent mode is a Raman-active mode. In MnWO$_4$, we find the highest $B_u$ mode at 767\,cm$^{-1}$
at 10\,K which is close in energy to the highest $B_g$ mode observed at 776\,cm$^{-1}$ at 5\,K.\cite{Iliev09}
Another possible origin of the asymmetric line shape is the mixing between transverse and longitudinal modes,
see Sec.\ \ref{SecLineShape}.
However, we may also adopt a phenomenological point of view, in which case we do not attempt to assign this silent mode to
a particular eigenmode of the compound but view it as a phenomenological source for an asymmetric line shape
of the infrared-active mode.
The dielectric function can be derived from the classical equations of motion for two damped harmonic oscillators
with eigenfrequencies $\omega_{\rm IR}$ and $\omega_{\rm s}$ and damping constants $\gamma_{\rm IR}$ and $\gamma_{\rm s}$, respectively,
\begin{eqnarray}
  \nonumber
  \ddot{x}_{\rm IR} &\! = \! & - \omega_{\rm IR}^2 x_{\rm IR} - \omega_{\rm IRs}^2 (x_{\rm IR}\! - \! x_{\rm s}) - \gamma_{\rm IR} \dot{x}_{\rm IR}
  + \frac{e E_0}{m} e^{- i \omega t}  \\
  \ddot{x}_{\rm s}  &\! = \! & - \omega_{\rm s}^2  x_{\rm s}  - \omega_{\rm IRs}^2 (x_{\rm s}\! - \! x_{\rm IR}) - \gamma_{\rm s}  \dot{x}_{\rm s}
  \label{coupling}
\end{eqnarray}
where $x_i$ denotes the displacement of oscillator $i \! \in \! \{{\rm IR},{\rm s}\}$, $t$ is the time,
$E_0$ the amplitude of the driving electric field,
$e$ and $m$ are the effective charge and mass of oscillator ``IR'', and $\omega_{\rm IRs}$ describes the coupling.
We are only interested in solutions for the polarization $P$
which oscillate with the frequency $\omega$ of the driving force,
\begin{equation}
P = \varepsilon_0 \left[\varepsilon(\omega) -1\right] E_0 e^{- i \omega t} = \frac{N}{V}\cdot e x_{\rm IR}
\end{equation}
with $N/V$ being the density, and $\varepsilon_0$ being the vacuum permittivity.
Thus the dielectric function reads
\begin{equation}
\label{silent}
 \varepsilon(\omega)  = 1 + \frac{\omega_{p,{\rm IR}}^2}{\omega_{\rm IR}^2 + \omega_{\rm IRs}^2 - \omega^2  - i \gamma_{\rm IR} \omega
                  -\frac{\omega_{\rm IRs}^4}{\omega_{\rm s}^2 + \omega_{\rm IRs}^2 - \omega^2  - i \gamma_{\rm s} \omega }}
\end{equation}
with the plasma frequency
\begin{equation}
\omega_{p,{\rm IR}}^2 = \frac{1}{\varepsilon_0} \frac{N}{V}\frac{e^2}{m} \, .
\end{equation}
Equation \ref{silent} reduces to the conventional Drude-Lorentz model (cf.\ Eq.\ \ref{EPSb}) for $\omega_{\rm IRs}\,=\,0$.
For finite $\omega_{\rm IRs}$, the damping is not a real function of $\omega$ but is effectively described by a complex term.
The main merit of this model is that it offers a Kramers-Kronig-consistent way for the description of a non-Lorentzian line shape,
which requires the introduction of three additional parameters ($\omega_{\rm s}$, $\gamma_{\rm s}$, and the
coupling $\omega_{\rm IRs}$).
Here, we use this asymmetric model only for the $B_u$ phonon mode with the highest eigenfrequency.
The total dielectric function thus reads
\begin{equation}\label{EPSacMonoclinicAsym}
    \begin{split}
        \hat{\varepsilon}_{ac}^{\rm \, asym} (\omega) &= \hat{\varepsilon}_{ac}^{\infty}
            + \sum_{i=1}^7 \frac{\omega_{p,i}^2}{\omega_{0,i}^2 - \omega^2
            - i \gamma_i \omega} \\
         & \qquad \qquad \times  S(\theta_i) \left(
            \begin{array}{cc}
              1  & 0\\
              0 & 0 \\
            \end{array}
         \right) S^{-1}(\theta_i) \\
         & \\
        & + \frac{\omega_{p,8}^2}{\omega_{0,8}^2 - \omega^2  - i \gamma_8  \omega
                  -\frac{\omega_{\rm IRs}^4}{\omega_{0,{\rm s}}^2 - \omega^2  - i \gamma_{\rm s} \omega }} \\
         & \qquad \qquad \times  S(\theta_8) \left(
            \begin{array}{cc}
              1  & 0\\
              0 & 0 \\
            \end{array}
         \right) S^{-1}(\theta_8) \,\, ,
  \end{split}
\end{equation}
with the abbreviations $\omega_{0,8}^2$\,:=\,$\omega_{\rm IR}^2 + \omega_{\rm IRs}^2$
and $\omega_{0,{\rm s}}^2$\,:=\,$\omega_{\rm s}^2 + \omega_{\rm IRs}^2$.

\subsection{Reflectance and Fitting Procedure}

For (near-)normal incidence, the reflectance $R_b (\omega)$ for $E \, \| \, b$ is given by
\begin{equation}
    R_b (\omega) = \left| \frac{1 - \sqrt{\varepsilon_b (\omega)}}{1 + \sqrt{\varepsilon_b (\omega)}}\right|^2  \,\, .
\end{equation}
whereas the reflectance $R_{ac} (\omega, \chi)$
is related to the tensor $\hat{\varepsilon}_{ac} (\omega)$ via\cite{Kuzmenko96,Kuzmenko01}
\begin{equation}\label{RacMonoclinic}
  \begin{split}
    R_{ac} \left( \omega, \chi \right) &= \left| \left( \left[\hat{1}
                    - \sqrt{\hat{\varepsilon}_{ac} (\omega)} \right]
                    \cdot \left[\hat{1} + \sqrt{\hat{\varepsilon}_{ac}
                    (\omega)} \right]^{-1} \right) \right.\\
    & \quad \times \left(
                    \begin{array}{c}
                      \cos \chi \\
                      \sin \chi \\
                    \end{array}
            \right) \bigg|^2 \,\, ,
  \end{split}
\end{equation}
where $\hat{1}$ and $(\dots)^{-1}$ denote the unity tensor and the inverse tensor, respectively.
The square root of the tensor $\hat{\varepsilon}_{ac}(\omega)$ is taken by rotating
$\hat{\varepsilon}_{ac}(\omega)$ to a diagonal form (employing a ``rotation'' matrix with complex entries),
then taking the square root for each matrix element, and the resulting matrix is rotated back
to its original basis.\cite{Kuzmenko96,Kuzmenko01}

The reflectivity $R_p (\omega,\alpha,\varphi)$ measured  for $p$-polarized light with (010) as plane of incidence
[see Sec.\ \ref{SecExp} and Fig.\ \ref{geometry}a)] is given by\cite{Koch74}
\begin{equation}\label{Rp}
    R_p (\omega,\alpha,\varphi) = \left| \frac{C\, \cos(\alpha) - \sqrt{\varepsilon_{ww}(\omega)-\sin^2(\alpha)}}
                                              {C\, \cos(\alpha) + \sqrt{\varepsilon_{ww}(\omega)-\sin^2(\alpha)}}\right|^2
\end{equation}
\begin{equation}
 \textrm{with} \,\,\,\,  C = \sqrt{\varepsilon_{uu}(\omega)\varepsilon_{ww}(\omega)-\varepsilon_{uw}^2(\omega)}  \,\, ,
\end{equation}
where $\alpha$\,=\,11$^\circ$ denotes the angle of incidence. Here, we use a Cartesian coordinate system $u$, $v$, $w$
with $v \, \| \, b$ and $w$ normal to the surface, and $\varphi$ denotes the angle between the $x$ axis
(which is fixed to the crystal axes) and the $u$ axis, i.e., the surface, see Fig.\ \ref{geometry}a).
In our case, $\varphi$\,=\,80$^\circ$.
In the basis $u$, $w$, the tensor $\hat{\varepsilon}_{ac}(\omega)$  is given by
\begin{equation}\label{EpsUWTensor}
\left(
      \begin{array}{cc}
        \varepsilon_{uu} & \varepsilon_{uw}\\
        \varepsilon_{uw} & \varepsilon_{ww}\\
       \end{array}
\right) =
    \hat{\varepsilon}_{ac}^{uw}(\varphi,\omega) = S(\varphi)\, \hat{\varepsilon}_{ac}(\omega) \, S^{-1}(\varphi)   \,\, ,
\end{equation}
with the rotation matrix $S(\varphi)$ as described in Eq.\ \ref{rotmat}.
For $E\! \perp \! b$, the parameters of the generalized Drude-Lorentz model and of the asymmetric oscillator model
were obtained by fitting the measured reflectance $R_{ac}(\omega,\chi)$ and $R_p(\omega,\alpha,\varphi)$ simultaneously.
In $R_p(\omega,11^\circ,80^\circ)$, transverse and longitudinal modes are strongly mixed.
The consideration of $R_p(\omega,11^\circ,80^\circ)$ thus offers an excellent test for the validity of the analysis,
see Sec.\ \ref{SecLineShape}.

For the fits we employed the MAGIX package (Modeling and Analysis Generic Interface for eXternal numerical codes)\cite{Moeller13},
which permits us to combine different optimization algorithms to make use of their specific advantages.
One may, e.g., combine a swarm algorithm to roughly localize a minimum in parameter space with
the Levenberg-Marquardt algorithm to optimize the parameters.
Due to the large number of parameters, we typically used only the Levenberg-Marquardt algorithm.
However, we employed the particle-swarm-optimization algorithm and the interval-nested-sampling algorithm\cite{Moeller13}
to search for alternative parameter values of weak oscillators.

\subsection{Kramers-Kronig-constrained variational approach}
\label{SecKK}

In the case of  a scalar dielectric function such as $\varepsilon_b(\omega)$,
a Kramers-Kronig analysis of, e.g., the normal-incidence reflectance $R_b(\omega)$
-- with appropriate extrapolations to $\omega$\,=\,0 and $\infty$ -- permits
a model-independent determination of $\varepsilon_b(\omega)$.
An extension to monoclinic symmetry has been discussed by Kuzmenko \emph{et al.}.\cite{Kuzmenko98}
Their approach is still based on a Kramers-Kronig analysis of the measured reflectance data
[in this case $R_{ac}(\omega,\chi)$] but employs a variational analysis.
More recently, a Kramers-Kronig-constrained generalization of the variational approach (KKvar)
has been discussed by Kuzmenko.\cite{Kuzmenko05}
In short, it uses a large number $N$ of oscillators, where $N$ is comparable to the number of measured data points.
The $N$ eigenfrequencies $\omega_1, \dots, \omega_N$ may, e.g., coincide with the frequency points of the
measured data. The fixed width of each oscillator is of the order of the step size $\omega_{i+1}\,-\,\omega_i$,
thus the contribution of any oscillator to the imaginary part Im$\{\varepsilon^{\rm KKvar}(\omega)\}$
is restricted to a small frequency interval.
The oscillator strengths of the $N$ oscillators are used to parametrize the frequency dependence of
Im$\{\varepsilon^{\rm KKvar}(\omega)\}$, whereas the real part Re$\{\varepsilon^{\rm KKvar}(\omega)\}$
is obtained via a Kramers-Kronig transformation.
Finally, the oscillator strengths are varied by fitting the experimental data.
Due to the large number of oscillators, this approach is well suited to describe asymmetric non-Lorentzian
line shapes of phonon modes or tiny spectral details and still yields a Kramers-Kronig-consistent result for the dielectric function.

Here, we use $3N$ oscillators with a triangular profile\cite{Kuzmenko05} at $N$ frequency points to parametrize
Im$\{\varepsilon_{xx}^{\rm KKvar}(\omega)\}$, Im$\{\varepsilon_{xz}^{\rm KKvar}(\omega)\}$,
and Im$\{\varepsilon_{zz}^{\rm KKvar}(\omega)\}$.
To reduce the calculational effort, we use a step size of 1\,cm$^{-1}$ with $\omega_1$\,=\,100\,cm$^{-1}$
and $\omega_N$\,=\,1000\,cm$^{-1}$.
A Kramers-Kronig analysis requires a reasonable extrapolation beyond the underlying frequency mesh
$\omega_1$ to $\omega_N$. Therefore, the result
$\hat{\varepsilon}_{ac}^{\rm \, gDL}$ of the generalized Drude-Lorentz fit (cf.\ Eq.\ \ref{EPSacMonoclinic})
is used as a starting point. The total dielectric function reads
\begin{equation}\label{epsilonTotal}
\hat{\varepsilon}_{ac}^{\rm total} (\omega) = \hat{\varepsilon}_{ac}^{\rm \, gDL} (\omega) + \hat{\varepsilon}_{ac}^{\rm KKvar} (\omega)  \, .
\end{equation}
The signs of Im$\{\varepsilon_{xx}^{\rm KKvar} (\omega)\}$, Im$\{\varepsilon_{xz}^{\rm KKvar} (\omega)\}$, and
Im$\{\varepsilon_{zz}^{\rm KKvar} (\omega)\}$ are arbitrary with the constraints Im$\{\varepsilon_{xx}^{\rm total} (\omega)\} \geq 0$
and Im$\{\varepsilon_{zz}^{\rm total} (\omega)\} \geq 0$.
We use this Kramers-Kronig-constrained variational approach to fit the reflectivity $R_{ac}(\omega,\chi)$
for three different polarization angles ($\chi$ = $0^\circ$, 30$^\circ$, and 90$^\circ$).

\section{Phonon modes}

Figures \ref{MnWO4RefFit010K} and \ref{MnWO4RefFit295K} show the reflectance $R_{ac}(\omega,\chi)$
for three different polarization angles $\chi$ as well as $R_p(\omega,\alpha,\varphi)$ and $R_b(\omega)$
at $T$\,=\,10\,K and 295\,K, respectively. Additionally, we show the reflectance as obtained from the
fits based on the generalized Drude-Lorentz model. We fitted $R_p(\omega,\alpha,\varphi)$ and $R_{ac}(\omega,\chi)$
for $\chi \in \{0^\circ, 30^\circ, 90^\circ\}$ simultaneously.
The fit parameters are listed in Tab.\ \ref{MnWO4FitTable}.
The asymmetric non-Lorentzian line shape observed in $R_{ac}(\omega,\chi)$ in case of the $B_u$ mode highest in energy
is discussed in Sec.\ \ref{SecLineShape}.
In Sec.\ \ref{SecT}, we address the temperature dependence of the spectra which is depicted in Fig.\ \ref{MnWO4RefallT}
for the low-frequency range.

\subsection{$A_u$ phonon modes}
\label{SecAu}

For $E \, \| \, b$, the agreement between data and fit is excellent, and the analysis is straightforward
(see bottom panels of Figs.\ \ref{MnWO4RefFit010K} and \ref{MnWO4RefFit295K}).
The spectra show seven $A_u$ modes, in agreement with the predictions of the factor-group analysis
for $T \! > \! T_{\rm N2} = 12.5$\,K.\@
We do not find any additional mode at 10\,K, i.e., below $T_{\rm N2}$.
The small discrepancies between data and fit around some of the maxima and minima of $R_b(\omega)$
can be attributed to small deviations from a Lorentzian line shape, typically caused by small
contributions stemming from the multi-phonon continuum.
Remarkably, the frequency of the highest $A_u$ mode amounts to 859\,cm$^{-1}$, which is unusually high
for a transition-metal oxide in which oxygen is the only light element.
However, comparable values have been reported for other tungstates $A$WO$_4$ with
divalent $A$\,=\,Cd, Ni, or Mg.\cite{Lacomba09,Daturi97,Kuzmin11,Ruiz10,Burcham98}
This mode can be assigned to a symmetric W-O(2) bond stretching phonon.
The high frequency reflects the strong bonding between the nominally hexa\-valent W ions and the O(2) ions,
as the shortest W-O(2) bond in MnWO$_4$ amounts to only 1.79\,\AA\ (see Fig.\ \ref{MnWO4Struc}).\cite{Macavei93}
Similar energies of stretching modes have been observed in, e.g., multiferroic Ni$_3$V$_2$O$_8$ and
$\alpha^\prime$-NaV$_2$O$_5$ with nominally penta- and tetravalent V ions.\cite{Yildirim08,Damascelli98}
The character of the other $A_u$ modes is given in Tab.\ \ref{MnWO4FitTable}.

\begin{table}[t]
    \centering
    \begin{tabular}{ccccccc}
        \hline  \hline
        \multicolumn{3}{c}{$B_u$ modes at 10\,K/295\,K} &   &   &   &  \\
        $\omega_0$ & $\omega_p$ & $\gamma$ & $\theta$ & $\Delta\varepsilon$ &  $\omega_0^{\rm calc}$[\onlinecite{Maczka11}] &   \\
        \hline
        139/137 &  310/288  &  0.7/ 2.6 &   22/ 20 & 4.98/4.44 & 163 &  \\
        201/197 &  393/386  &  1.4/ 5.8 &  162/157 & 3.84/3.84 & 186 &  \\
        241/239 &  424/390  &  0.5/ 1.1 &  121/122 & 3.10/2.66 & 206 &  \\
        273/277 &  865/822  &  2.7/13.2 &   72/ 72 & 10.1/8.84 & 263 &  \\
        283/283 &   65/ 45  &  3.0/ 4.9 &  168/167 & 0.05/0.02 & 323 &  \\
        455/453 &  345/319  &  5.9/14.8 &    0/  1 & 0.58/0.50 & 467 &  \\
        554/553 & 1102/1071 &  8.3/18.6 &  123/123 & 3.96/3.76 & 576 &  \\
        767/771 & 1043/1008 &  6.6/12.4 &   34/ 34 & 1.85/1.71 & 777 &  \\
        \hline
        \hline
         &  &  &  &  &  &  \\
        \multicolumn{3}{c}{$A_u$ modes at 10\,K/295\,K} &   &   &  & \\
        $\omega_0$ & $\omega_p$ & $\gamma$ & $\Delta\varepsilon$ &  char.[\onlinecite{Maczka11}] &  $\omega_0^{\rm calc}$[\onlinecite{Maczka11}] &  \\
        \hline
        174/168 & 348/342 &  1.1/ 5.4 & 3.99/4.12   &  T$^\prime$                    &  156 & \\
        309/306 & 133/110 &  2.3/ 4.2 & 0.18/0.13   &  $\tau$ + $\delta_{sc}$        &  246 & \\
        341/341 & 501/494 &  1.5/ 4.2 & 2.15/2.11   &  $\delta_{sc}$ + $\tau$        &  410 & \\
        419/416 & 308/288 &  5.2/11.5 & 0.54/0.48   &  $\delta_{as}$ + $\tau$        &  455 & \\
        500/498 & 513/518 &  9.2/21.0 & 1.05/1.08   &  $\nu_{as}$ + $\delta_{sc}$    &  547 & \\
        664/663 & 756/759 & 11.5/21.8 & 1.30/1.31   &  $\nu_{as}$                    &  671 & \\
        859/860 & 425/422 &  6.2/10.0 & 0.24/0.24   &  $\nu_{s}$                     &  837 & \\
        \hline
        \hline
    \end{tabular}
\caption{Parameters of the generalized Drude-Lorentz model for $T$\,=\,10\,K and 295\,K.\@
Here, $\Delta \varepsilon$\,=\,$(\omega_p/\omega_0)^2$ denotes the oscillator strength.
The parameters $\omega_0$, $\omega_{p}$, and $\gamma$ are given in units of [cm$^{-1}$],
whereas the angle $\theta$ is given in $[^{\circ}]$. For the high-frequency dielectric constant
at $T$ = 10\,K (295\,K) we find
$\varepsilon_{xx}^{\infty}$ = 5.59 (5.25),
$\varepsilon_{yy}^{\infty}$ = 5.48 (5.35),
$\varepsilon_{zz}^{\infty}$ = 6.25 (5.88),
and $\varepsilon_{xz}^{\infty}$ = 0.29 (0.25).
The calculated values for $\omega_0$ (right column) and the characters of the $A_u$ modes are
reproduced from Ref.\ [\onlinecite{Maczka11}], where T$^\prime$ stands for T$^\prime$(Mn)+T$^\prime$(W) and
denotes a lattice translational mode,
$\tau$ is a W-O(2) twisting mode, $\delta_{sc}$ a W-O(2) scissoring mode, $\delta_{as}$ an antisymmetric
W-O(1) bending mode, and $\nu_{as}$ [$\nu_s$] an antisymmetric W-O(1) [symmetric W-O(2)] stretching mode.\cite{Maczka11}
}
\label{MnWO4FitTable}
\end{table}

As far as the eigenfrequencies $\omega_{0,i}$ of the $A_u$ modes along the unique $b$ axis are concerned,
the results of Refs.\ [\onlinecite{Choi10,Maczka11}] and our data agree very well with each other.
With the exception of the lowest $A_u$ mode, the values for $\omega_{0,i}$ agree within about 1-2\,\%.
In MnWO$_4$, we and Choi {\emph et al.}\cite{Choi10} found the lowest mode at 168\,cm$^{-1}$ at room temperature,
whereas it was reported at 180 and 182\,cm$^{-1}$ in Mn$_{0.85}$Co$_{0.15}$WO$_4$ and Mn$_{0.97}$Fe$_{0.03}$WO$_4$,
respectively.\cite{Maczka11}
Note that the values of the damping $\gamma_i$ and the oscillator strength $\Delta\varepsilon_i$ are not reported
for the single-crystal data in Refs.\ [\onlinecite{Choi10,Maczka11}].

The $A_u$ phonon parameters enable us to determine the contribution
Re$\{\varepsilon_b^{\rm high}\} = \varepsilon_{yy}^\infty + \Sigma_{i=1}^7 \Delta\varepsilon_i$
of phonons and of excitations at higher energies to the quasi-static dielectric constant
Re$\{\varepsilon_b\}$, i.e., for frequencies well below the phonon range.
Combining this result with low-frequency data measured
at 96.8\,kHz and 45\,MHz -- far below the frequency range of a possible electromagnon --
allows us to estimate the contribution of a possible electromagnon to Re$\{\varepsilon_b\}$,
which will be discussed together with the temperature dependence in Sec.\ \ref{SecT}.
At 10\,K we find Re$\{\varepsilon_b^{\rm high}\}$\,=\,14.9. This has to be compared to the results reported
from impedance measurements using LCR meters, Re$\{\varepsilon_b\}$\,=\,16.4 at 1\,MHz (Ref.\ \onlinecite{Arkenbout06})
and 12.3 at 1\,kHz (Ref.\ \onlinecite{Taniguchi06}). Note that impedance measurements typically show a high relative
accuracy but larger errors concerning the absolute value due to uncertainties in the size and shape of the
electrodes.

\subsection{$B_u$ phonon modes}
\label{SecBu}

\begin{figure}[t]
\includegraphics[width=0.96\columnwidth,clip]{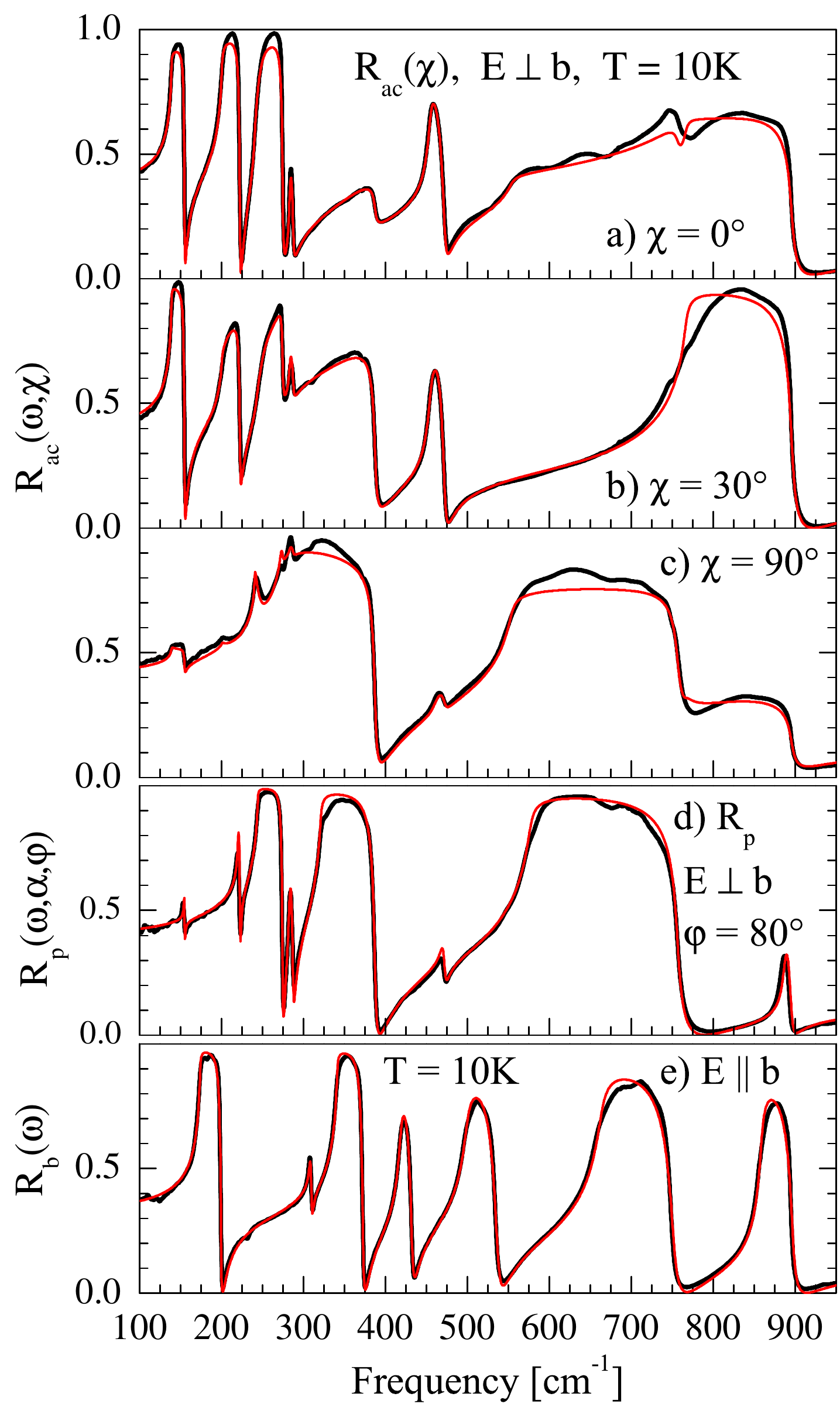}
\caption{(Color online) Reflectance of MnWO$_4$ at $T$\,=\,10\,K (black)
and generalized Drude-Lorentz fit (red).
a)\,-\,c) $R_{ac}(\omega,\chi)$ for different polarization angles $\chi$ as defined in Fig.\ \ref{MnWO4RefAllAngles}.
d) Reflectance $R_p(\omega,11^\circ,80^\circ)$ for $p$-polarized light incident within the (010) plane
with $E \! \perp \! b$, see
Fig.\ \ref{geometry}a) and Eq.\ \ref{Rp}.
e) Reflectance $R_b(\omega)$ for $E \, \| \, b$. }
\label{MnWO4RefFit010K}
\end{figure}

\begin{figure}[t]
\includegraphics[width=0.96\columnwidth,clip]{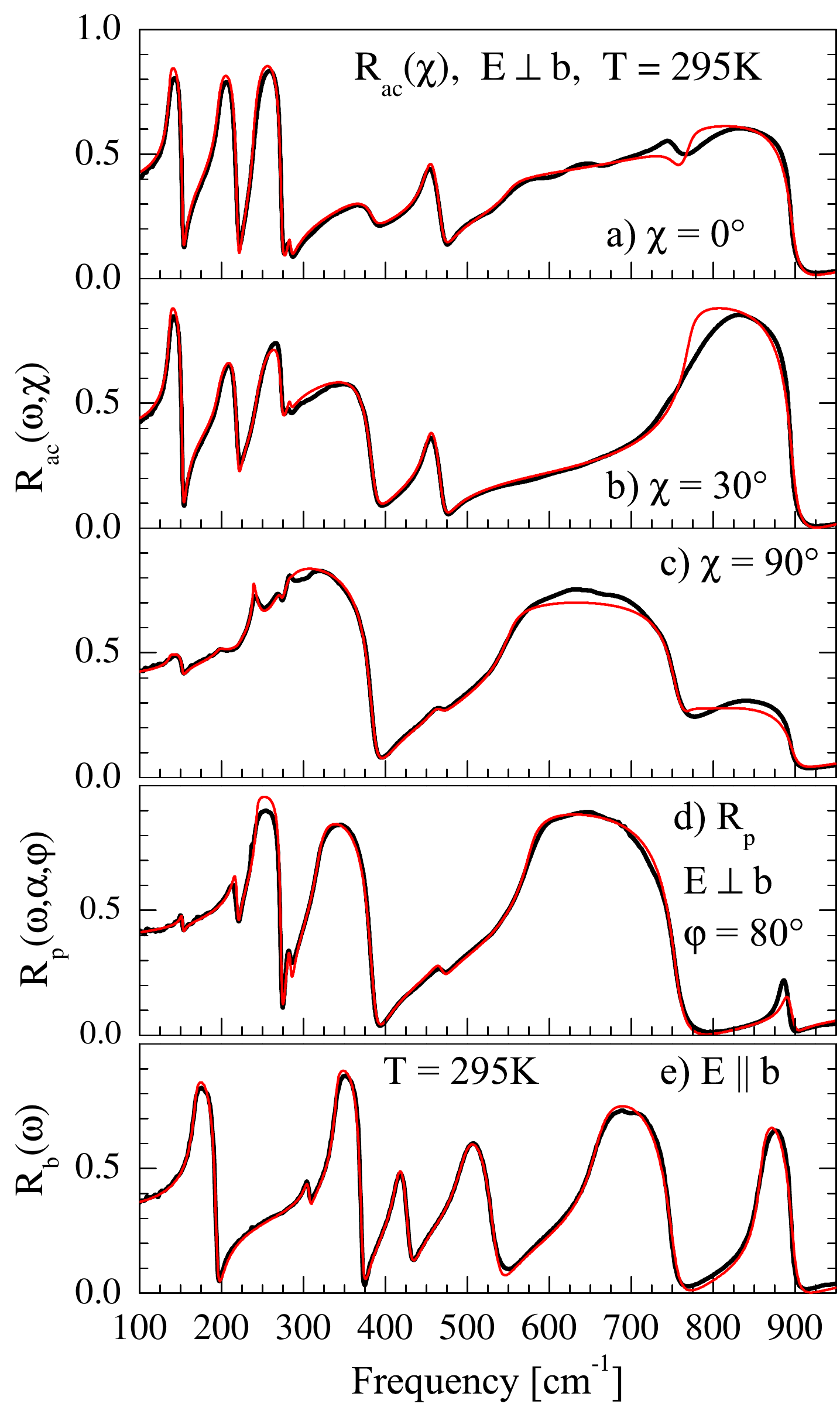}
\caption{(Color online) Reflectance of MnWO$_4$ at $T$\,=\,295\,K (black) and generalized Drude-Lorentz fit (red).
a)\,-\,c) $R_{ac}(\omega,\chi)$ for different polarization angles $\chi$,
d) $R_p(\omega,11^\circ,80^\circ)$,
e) $R_b(\omega)$.
}
\label{MnWO4RefFit295K}
\end{figure}

\begin{figure}[t!]
\includegraphics[width=0.9\columnwidth,clip]{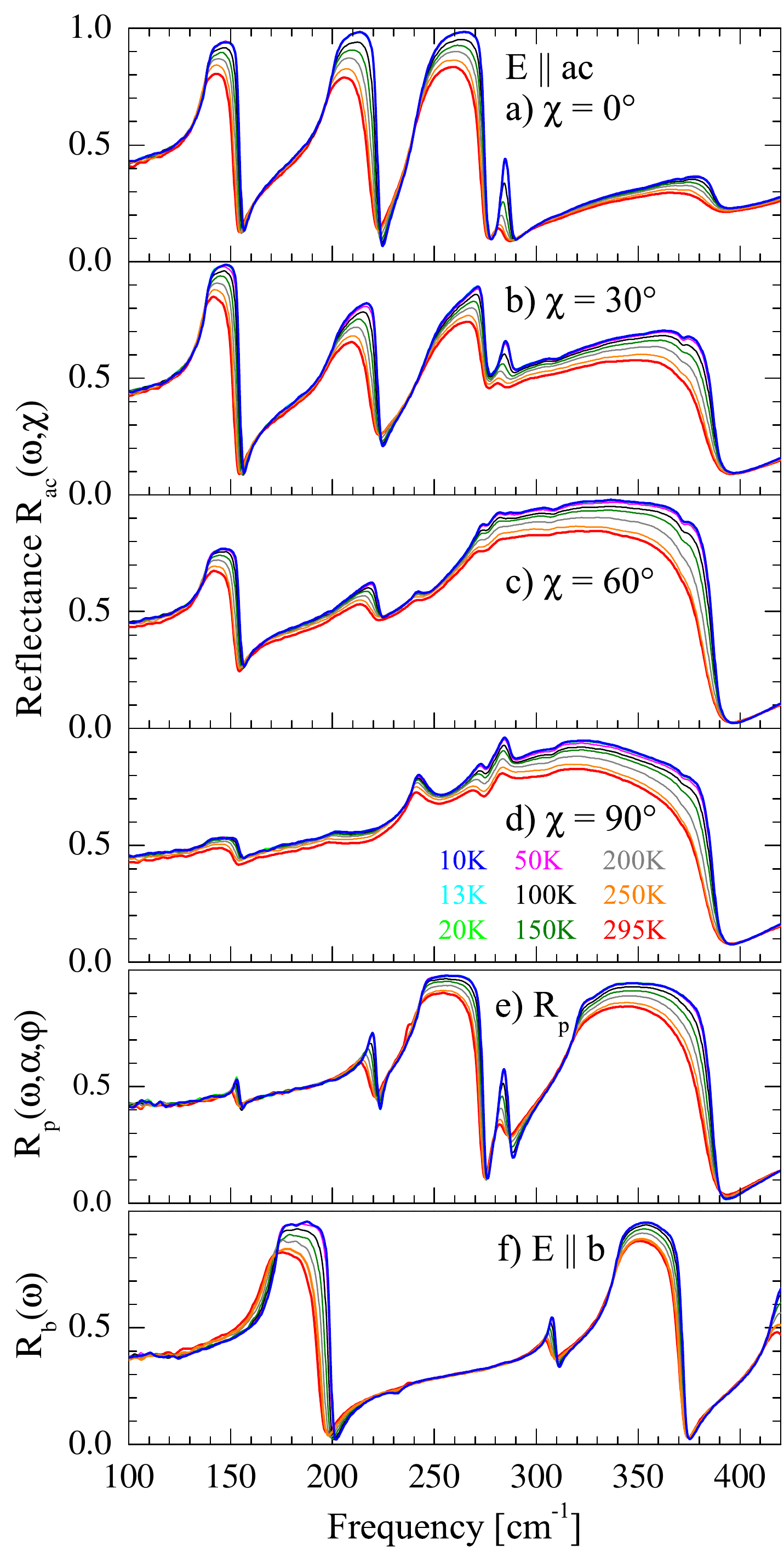}
\caption{(Color online) Temperature dependence of the reflectance in the low-frequency range.
Panels a)-d): $R_{ac}(\omega,\chi)$, e) $R_p(\omega, 11^\circ, 80^\circ)$, f) $R_b(\omega)$.
The largest changes of the spectral weight are observed for the weak $B_u$ mode at 283\,cm$^{-1}$
and for the weak $A_u$ mode at 309\,cm$^{-1}$. The lowest $A_u$ mode shows the largest softening. }
\label{MnWO4RefallT}
\end{figure}

The case of $E\! \perp \! b$ requires a more careful analysis. The interplay of partially overlapping modes
with different rotation angles $\theta_i$ gives rise to complex line shapes in $R_{ac}(\omega,\chi)$,\cite{Ivanovski07}
see Figs.\ \ref{MnWO4RefFit010K} -- \ref{MnWO4RefallT}.
Therefore it is more difficult to disentangle the contributions of the different modes.
We find that the data can be described by a sum of eight $B_u$ modes, as predicted by the factor-group
analysis for $T \! > \! 12.5$\,K (cf.\ Eq.\ \ref{AuBu}).
Eight separate $B_u$ modes are most easily recognizable in the data of $R_p(\omega)$. However, the eigenfrequencies
can be inferred more easily from $R_{ac}(\omega,\chi)$.
Five of the eight $B_u$ modes have an eigenfrequency lower than 300\,cm$^{-1}$, see Tab.\ \ref{MnWO4FitTable}.
Four out of these five modes are easily recognized in the spectrum of $R_{ac}(\omega)$ for $\chi$\,=\,0$^\circ$
below 300\,cm$^{-1}$ at $T$\,=\,10\,K (see Fig.\ \ref{MnWO4RefallT}).
These four modes have eigenfrequencies of
$\omega_{0,1}$\,=\,139\,cm$^{-1}$, $\omega_{0,2}$\,=\,201\,cm$^{-1}$,
$\omega_{0,3}$\,=\,241\,cm$^{-1}$, and $\omega_{0,5}$\,=\,283\,cm$^{-1}$.
The mode with $\omega_{0,4}$\,=\,273\,cm$^{-1}$ has a much larger oscillator strength, giving rise to the
pronounced Reststrahlen band observed between about 250\,cm$^{-1}$ and 400\,cm$^{-1}$ for $\chi$\,=\,90$^\circ$.
The eigenfrequency $\omega_{0,4}$\,=\,273\,cm$^{-1}$ corresponds to the low-frequency edge of the Reststrahlen band,
whereas the steep drop of $R_{ac}(\omega,90^\circ)$ at about 390\,cm$^{-1}$ can be identified with its longitudinal
eigenfrequency (see Secs.\ \ref{SecDielec} and \ref{SecLineShape} for a discussion of the mixing of LO and TO modes).
This mode with $\theta_4$\,=\,72$^\circ$ gives rise to the peculiar feature peaking at about 375\,cm$^{-1}$
for $\chi$\,=\,0$^\circ$ and 30$^\circ$. The sixth mode at $\omega_{0,6}$\,=\,455\,cm$^{-1}$ is well separated
in frequency from all other modes and thus can be observed easily for all values of $\chi$ considered here.
Due to the strong orientational dispersion of $\hat{\varepsilon}_{ac}(\omega)$
in the vicinity of this mode (see below), it also affects $R_{ac}(\omega,90^\circ)$
although $\chi$\,=\,90$^\circ$ is almost orthogonal to its orientation $\theta_6 \approx 0^\circ$.


Modes number seven and eight determine $R_{ac}(\omega)$ between 500 and 900\,cm$^{-1}$.
With $\theta_7$\,=\,123$^\circ$ and $\theta_8$\,=\,34$^\circ$, these two modes are nearly orthogonal to each
other. Accordingly, $R_{ac}(\omega,30^\circ)$ predominantly shows the higher mode with $\omega_{0,8}$\,=\,767\,cm$^{-1}$.
The signature of the lower mode will be most pronounced for $\chi \approx \theta_7$\,=\,123$^\circ$, whereas the
measured data sets with $\chi$\,=\,0$^\circ$, $60^\circ$, and 90$^\circ$ show complicated line shapes which
reflect the existence of both modes.
Note that the pronounced peak at about 750\,cm$^{-1}$ for $\chi$\,=\,0$^\circ$ does not require to invoke
a further infrared-active mode. This peak is located at the frequency of the steep drop of $R_{ac}(\omega,90^\circ)$,
i.e., at the longitudinal eigenfrequency of mode seven with $\omega_{0,7}$\,=\,554\,cm$^{-1}$
(see also Fig.\ \ref{MnWO4RefAllAngles}).
For comparison, it is instructive to consider the nearly triangular hump around 375\,cm$^{-1}$
for $\chi$\,=\,0$^\circ$ which stems from the phonon mode with $\theta_4$\,=\,72$^\circ$.
Similarly, the peculiar shape of the peak at 750\,cm$^{-1}$ for $\chi$\,=\,0$^\circ$ or 60$^\circ$ (see Figs.\
\ref{MnWO4RefAllAngles} and \ref{MnWO4Ref60deg}) originates from the phonon mode with $\theta_7$\,=\,123$^\circ$,
i.e., roughly $\pm 60^\circ$ different from the value of $\chi$.
Moreover, lattice dynamical calculations\cite{Maczka11} for MnWO$_4$ predict only two $B_u$ modes
above 500\,cm$^{-1}$, namely at 576\,cm$^{-1}$ and 777\,cm$^{-1}$, in reasonable agreement with our
experimental values of 554\,cm$^{-1}$ and 767\,cm$^{-1}$.
Also first-principles calculations\cite{Kuzmin11} for NiWO$_4$ find only two modes with $B_u$
symmetry above 500\,cm$^{-1}$.

The existence of eight $B_u$ modes is corroborated by the data for $R_p(\omega,\alpha,\varphi)$ measured with
$p$-polarized light and (010) as plane of incidence [see panels (d) of Figs.\ \ref{MnWO4RefFit010K} and
\ref{MnWO4RefFit295K} and Eq.\ \ref{Rp}].
In particular, $R_p$ shows the five lowest modes clearly separated from each other and only two $B_u$ modes above 500\,cm$^{-1}$.
For this measurement geometry, the modes show a strong LO-TO mixing, and the character changes from predominantly
transverse to predominantly longitudinal as a function of the orientation angle $\theta_i$ of the mode with respect
to the wave vector $k$. Accordingly, both the apparent peak position and the oscillator strength depend
strongly on the angle of incidence $\alpha$ and on the angle $\varphi$, which describes the orientation of
the $ac$ plane with respect to the surface [see Fig.\ \ref{geometry}a) and Eq.\ \ref{Rp}].
This explains in particular the pronounced changes between $R_{ac}(\omega)$ and $R_p(\omega)$ observed for
modes 4 and 8: $R_p(\omega,11^\circ,80^\circ)$
shows the high-frequency edges of the phonon modes (i.e., the LO frequencies) at the same frequencies
as $R_{ac}(\omega,\chi)$, but the apparent oscillator strength is very different in $R_p$ and $R_{ac}$.
As a result, all eight $B_u$ modes are clearly separated from each other in $R_p$, but the apparent order of modes 4 and 5
is reversed. The weak mode 5 gives rise to a feature close to $\omega_{0,5}$\,=\,283\,cm$^{-1}$ also in $R_p$,
but the much stronger mode 4 with $\omega_{0,4}$\,=\,273\,cm$^{-1}$ appears as a band between 300\,cm$^{-1}$
and 400\,cm$^{-1}$ in $R_p$.
With small exceptions, our fit describes both $R_p(\omega,\alpha,\varphi)$ and $R_{ac}(\omega,\chi)$ very well.
This clearly demonstrates that $\hat{\varepsilon}_{ac}(\omega)$ has been determined correctly (cf.\ Sec. \ref{SecLineShape}).

\begin{figure}[t!]
\includegraphics[width=0.9\columnwidth,clip]{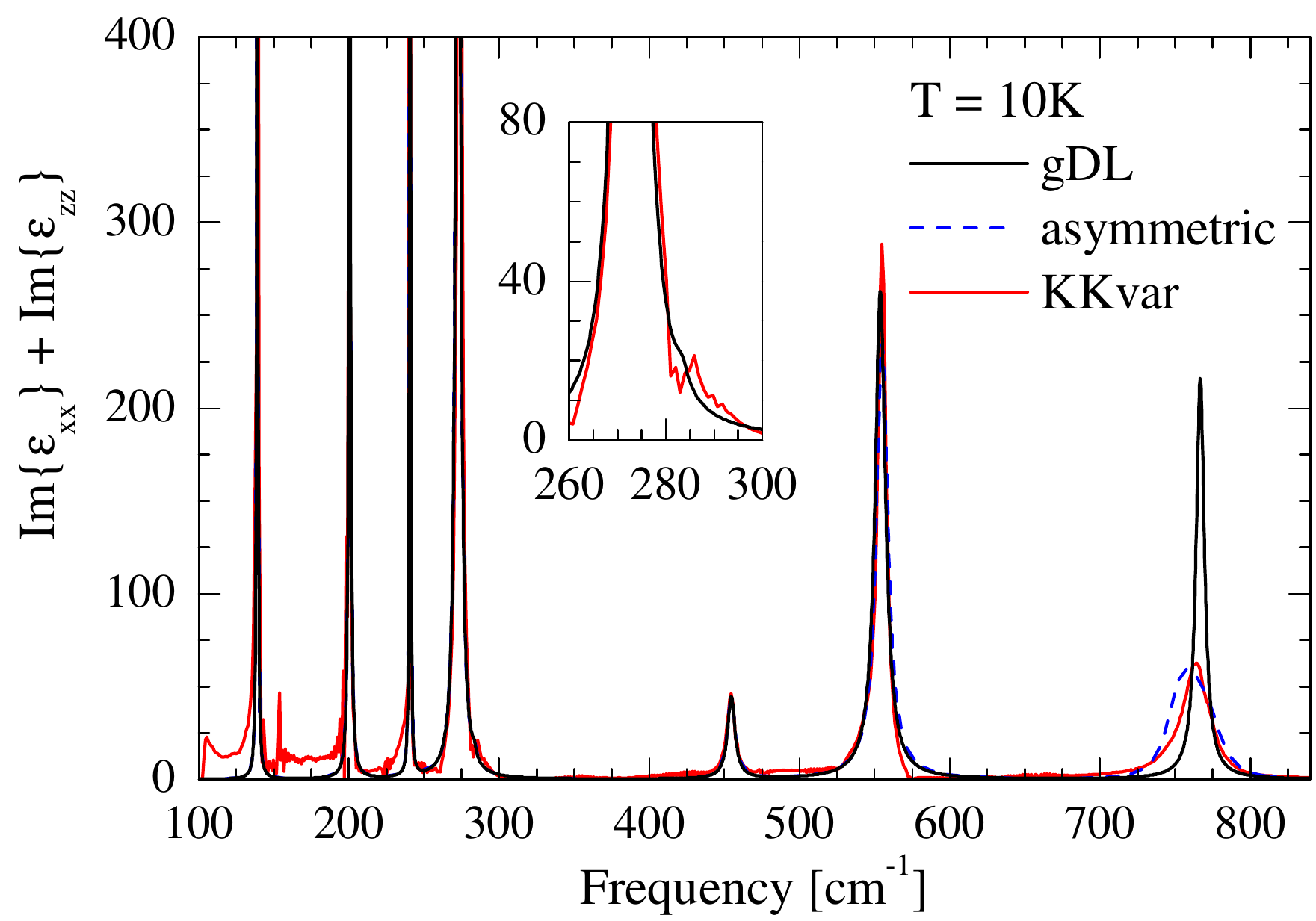}
\caption{(Color online) The sum Im$\{\varepsilon_{xx}(\omega)\}$\,+\,Im$\{\varepsilon_{zz}(\omega)\}$ shows all
eight $B_u$ phonon modes.
A Kramers-Kronig-constrained variational analysis of $R_{ac}(\omega,\chi)$ (KKvar, red line, cf.\ Sec.\ \ref{SecKK})
confirms the results of a generalized Drude-Lorentz fit (gDL,  black line, cf.\ Eq.\ \ref{EPSacMonoclinic})
with the exception of the line shape of the highest mode at 767\,cm$^{-1}$. For this mode,
the KKvar result supports a non-Lorentzian line shape as described by
the asymmetric oscillator model (blue dashed line, cf.\ Eq.\ \ref{EPSacMonoclinicAsym}).
Inset: weak mode at 283\,cm$^{-1}$ on an enlarged scale.
}
\label{MnWO4eps2KKfit}
\end{figure}

\begin{figure}[t]
\includegraphics[width=0.9\columnwidth]{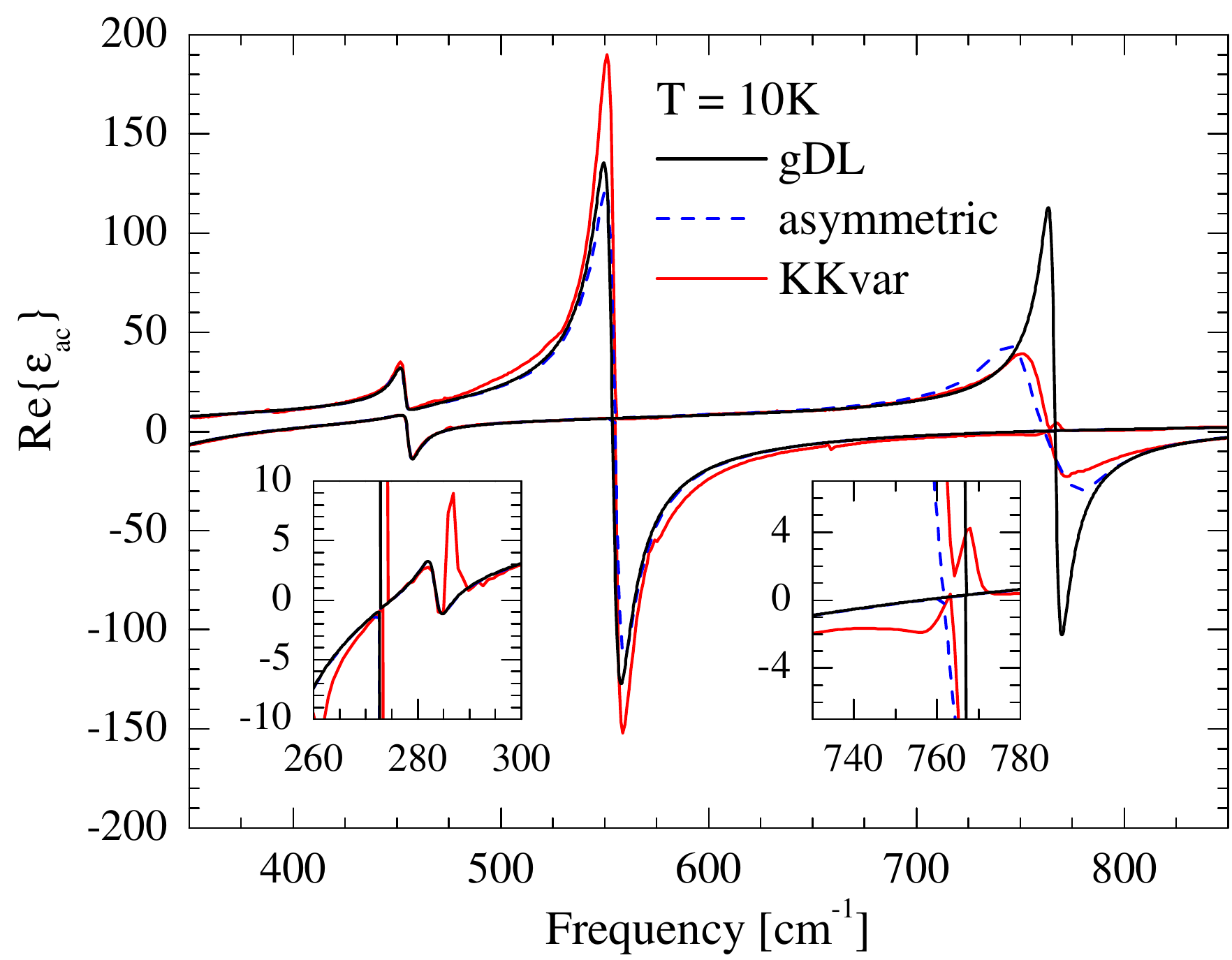}
\caption{(Color online) The two entries of the diagonal form of the real part Re$\{\hat{\varepsilon}_{ac}(\omega,\chi)\}$
at 10\,K as obtained from a generalized Drude-Lorentz fit (gDL,  black lines, cf.\ Eq.\ \ref{EPSacMonoclinic}),
a Kramers-Kronig-constrained variational analysis of $R_{ac}(\omega,\chi)$ (KKvar, red lines, cf.\ Sec.\ \ref{SecKK}),
and a fit using an asymmetric, non-Lorentzian line shape for the highest $B_u$ mode (blue dashed lines, cf.\ Eq.\ \ref{EPSacMonoclinicAsym}).
Insets: same data on an enlarged scale.
}
\label{MnWO4Epsilon}
\end{figure}

A Kramers-Kronig-constrained variational analysis of $R_{ac}(\omega,\chi)$ for $\chi \in \{0^\circ, 30^\circ, 90^\circ\}$
(see Sec.\ \ref{SecKK}) supports our results from the generalized Drude-Lorentz model for
the number of modes and for the properties of the lower seven modes.
Figure \ref{MnWO4eps2KKfit} compares Im$\{\varepsilon_{xx}(\omega)\}$\,+\, Im$\{\varepsilon_{zz}(\omega)\}$ obtained
by the two approaches.
The Kramers-Kronig-constrained variational analysis equally shows four strong $B_u$ modes below 300\,cm$^{-1}$,
a weak feature that corresponds to the mode at $\omega_{0,5}$\,=\,283\,cm$^{-1}$ (see insets of Figs.\ \ref{MnWO4eps2KKfit}
and \ref{MnWO4Epsilon}), one mode at about 450\,cm$^{-1}$, and two modes above 500\,cm$^{-1}$.
The Kramers-Kronig-constrained variational analysis employs a fixed line width and thus encounters problems
to precisely describe the phonon modes number 1 and 3 with line widths smaller than 1\,cm$^{-1}$. This explains the spike
observed at about 154\,cm$^{-1}$ in Fig.\ \ref{MnWO4eps2KKfit}.

The results of the two approaches for the diagonalized form of the real part of the dielectric function
Re$\{\hat{\varepsilon}_{ac}\}$ are given in Fig.\ \ref{MnWO4Epsilon} for the high-frequency range.
The main discrepancy between the generalized Drude-Lorentz model and the Kramers-Kronig-constrained variational approach
is observed for the line shape of the highest $B_u$ mode (see Sec.\ \ref{SecLineShape}).

\begin{figure}[bt]
\includegraphics[width=0.9\columnwidth,clip]{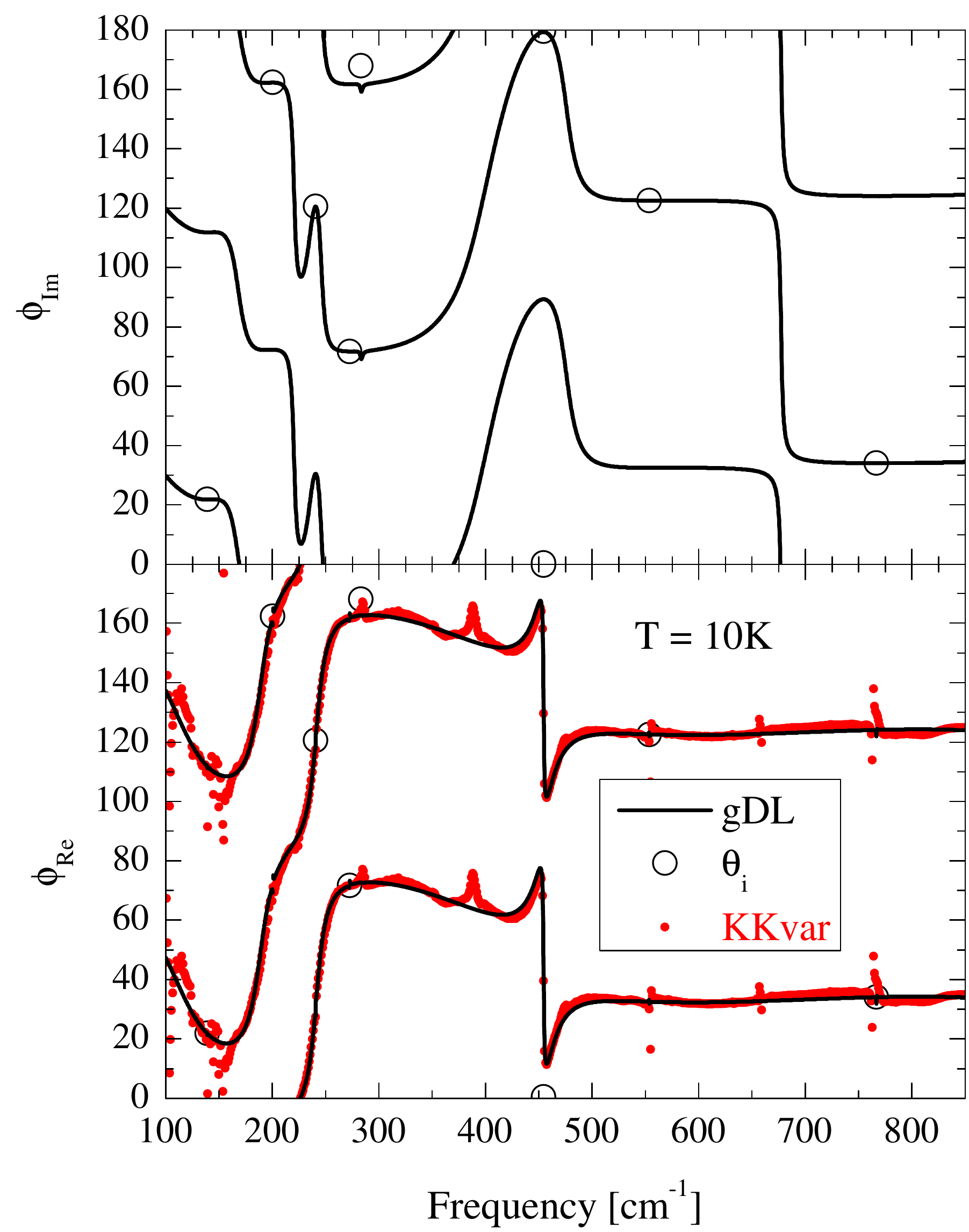}
\caption{The rotation angles $\phi_{\rm Im}(\omega)$ of Im$\{\hat{\varepsilon}_{ac}(\omega,\chi)\}$ (solid lines, top panel)
and $\phi_{\rm Re}(\omega)$ of Re$\{\hat{\varepsilon}_{ac}(\omega,\chi)\}$ (solid lines, bottom panel)
at 10\,K as obtained from a generalized Drude-Lorentz fit (gDL,  cf.\ Eq.\ \ref{EPSacMonoclinic}).
Full red symbols: $\phi_{\rm Re}(\omega)$ as obtained from a Kramers-Kronig-constrained variational analysis.
The angles $\phi_{\rm Im}(\omega)$ and $\phi_{\rm Re}(\omega)$ are plotted modulo 90$^\circ$.
Open symbols: The angle $\theta_i$ between the dipole moment of mode $i$ and the $x$ axis (cf.\ Tab.\ II). }
\label{MnWO4Angles}
\end{figure}

The rotation angles $\phi_{\rm Im}(\omega)$, $\phi_{\rm Re}(\omega)$, and the angles $\theta_i$ of the different oscillators
of the generalized Drude-Lorentz model are plotted in Fig.\ \ref{MnWO4Angles}.
Below 500\,cm$^{-1}$, MnWO$_4$ shows a pronounced orientational dispersion of $\hat{\varepsilon}_{ac}(\omega)$.
The two high-energy modes above 500\,cm$^{-1}$ originate from stretching/bending vibrations of W-O bonds of
the tightly bound WO$_6$ octahedra.
Remarkably, the rotation angles $\theta_7$\,=\,123$^\circ$ and $\theta_8$\,=\,34$^\circ$ of these two modes
agree very well with the projections of the W-O bonds on the $ac$ plane, see Fig.\ \ref{MnWO4Struc}.
The O(2)-O(2) edge of the WO$_6$ octahedra is rotated by about 30$^\circ$ with respect to the $x$ axis,
and the rotation of the shorter O(1)-W-O(1) bond (light red bonds in Fig.\ \ref{MnWO4Struc})
amounts to about 123$^\circ$.

In total, we have identified both the seven $A_u$ modes and the eight $B_u$ modes predicted by
a factor-group analysis. This has been claimed before both by Choi \emph{et al.}\cite{Choi10}
and by Maczka \emph{et al.},\cite{Maczka11} but both studies investigated only two polarization
directions within the $ac$ plane. Our analysis demonstrates that this is clearly not sufficient
to determine the modes with $B_u$ symmetry in this mono\-clinic compound.
In the analysis of the single-crystal data of Maczka \emph{et al.},\cite{Maczka11} modes are missing
at 273\,cm$^{-1}$, 283\,cm$^{-1}$, and 554\,cm$^{-1}$ in Mn$_{0.85}$Co$_{0.15}$WO$_4$ and
at 201\,cm$^{-1}$, 273\,cm$^{-1}$, and 455\,cm$^{-1}$ in Mn$_{0.97}$Fe$_{0.03}$WO$_4$.
Partially, these modes have been observed in the polycrystalline samples.\cite{Maczka11}
Choi {\emph et al.}\cite{Choi10} reported 8 frequencies for $E \, \| \, a$ and 6 more
frequencies for $E \, \| \, c$, 5 (4) of them nearly degenerate with the ones reported for
$E \, \| \, a$ ($b$). Our data reveal that several of these modes are actually $A_u$ modes,
whereas the $B_u$ modes at 273\,cm$^{-1}$, 283\,cm$^{-1}$, 554\,cm$^{-1}$, and 767\,cm$^{-1}$
are missing. At first sight, it
may seem surprising that in particular the modes with large oscillator strength at 273\,cm$^{-1}$
and at 554\,cm$^{-1}$ have been overlooked.
However, our reflectivity data show that in particular the line shapes of the strong modes with
large LO-TO splitting -- giving rise to spectral overlap with other bands\cite{Ivanovski07} --
depend strongly on the angle $\chi$. Moreover, the reflectivity spectrum of the $B_u$ modes
strongly depends on the measurement geometry, i.e., on the direction of the wavevector.
Weaker modes give rise to sharper features, which facilitates the determination of their eigenfrequencies.

\subsection{Line shape of the highest $B_u$ mode}
\label{SecLineShape}

\begin{figure}[t]
\includegraphics[width=0.85\columnwidth]{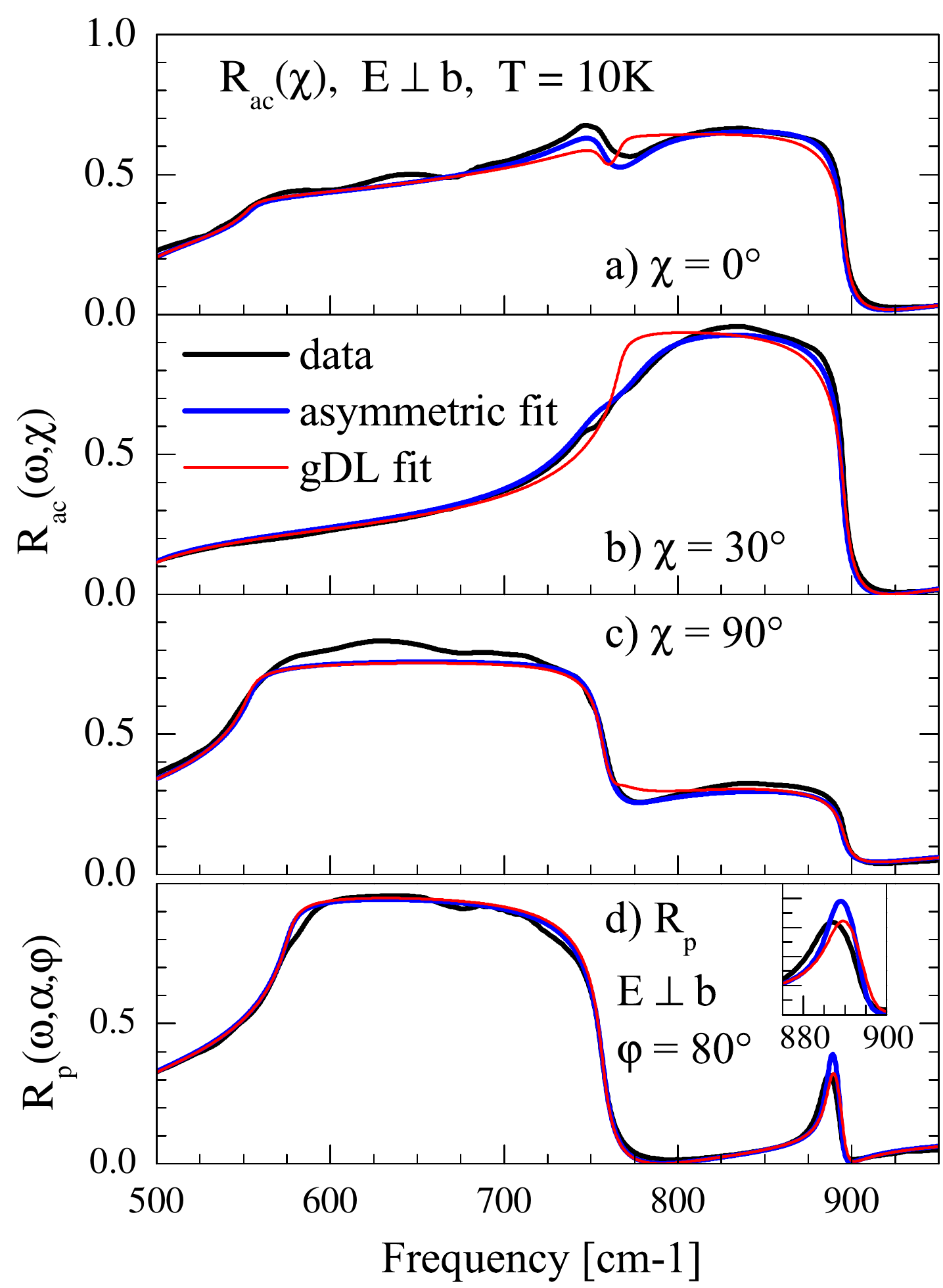}
\caption{(Color online) Comparison of fits based on the generalized Drude-Lorentz model (red, cf.\ Eq.\ \ref{EPSacMonoclinic})
and the asymmetric oscillator model (blue, cf.\ Eq.\ \ref{EPSacMonoclinicAsym}).
In both cases, $R_p(\omega)$ and $R_{ac}(\omega,\chi)$ with $\chi \in \{0^\circ, 30^\circ, 90^\circ\}$ were fitted simultaneously.
Black: measured reflectance data.
Inset: highest $B_u$ mode on an enlarged scale.  }
\label{FigBu7}
\end{figure}

The line shape of $R_{ac}(\omega,\chi)$ is not described very well between about 700\,cm$^{-1}$ and 800\,cm$^{-1}$
if we stick to eight modes with a Lorentzian line shape (see Fig.\ \ref{FigBu7}).
We emphasize that adding a further Lorentzian mode to $\hat{\varepsilon}_{ac}^{\rm \, gDL}(\omega)$
in Eq.\ \ref{EPSacMonoclinic} does not significantly improve the quality of the fit.
The steep drop of the Reststrahlen band at about 900\,cm$^{-1}$ indicates a small value of the damping $\gamma_8$
of the eighth $B_u$ mode, whereas the much more gradual rise at the low-frequency side of this mode
between 700\,cm$^{-1}$ and 800\,cm$^{-1}$ for $\chi$\,=\,30$^\circ \approx \theta_8$ is a clear signature of a larger damping.
Accordingly, a fit based on Eq.\ \ref{EPSacMonoclinicAsym} with an asymmetric line shape for the mode with
$\omega_{0,8}$\,=\,767\,cm$^{-1}$ yields a much better description of $R_{ac}(\omega,\chi)$ (cf.\ blue lines
in Fig.\ \ref{FigBu7}). The fit parameters are given in Table \ref{MnWO4FitTableAsym}.
An asymmetric line shape is supported by the Kramers-Kronig-consistent variational analysis of $R_{ac}(\omega,\chi)$
(see Figs.\ \ref{MnWO4eps2KKfit} and \ref{MnWO4Epsilon}).

\begin{table}[t]
    \centering
    \begin{tabular}{ccccccccc}
        $\omega_{0,8}$ & $\omega_{p,8}$ & $\gamma_8$ & $\theta_8$ & \hspace*{2mm} &
        $\omega_{\rm 0,s}$ & $\gamma_{\rm s}$ & \hspace*{2mm} & $\omega_{\rm IRs}$ \\
        \hline
        762 & 1048 &  3.9 &   33 & & 765 &  162  & &  56   \\
    \end{tabular}
\caption{Parameters of the asymmetric oscillator model for the eighth $B_u$ mode at $T$\,=\,10\,K.\@
}
\label{MnWO4FitTableAsym}
\end{table}

In contrast to $R_{ac}(\omega,\chi)$, $R_p(\omega)$ is described very well by the generalized Drude-Lorentz model
[red line in panel d) of Fig.\ \ref{FigBu7}]. In fact, the description of the highest $B_u$ mode in $R_p(\omega)$ becomes
slightly worse if we use the asymmetric oscillator model, which overestimates the absolute value of $R_p(\omega)$ at the
maximum of the highest mode at about 880\,cm$^{-1}$.
This is a compromise of the fit which aims at a simultaneous description of $R_{ac}(\omega,\chi)$ and $R_p(\omega)$.
Note that the agreement between $R_{ac}(\omega,\chi)$ and the fit is not improved significantly if we fit only
$R_{ac}(\omega,\chi)$ but not $R_p(\omega)$.

One possible source for the asymmetric line shape in $R_{ac}(\omega,\chi)$ is a mixing of transverse and longitudinal modes.
Our analysis of $R_{ac}(\omega,\chi)$ assumes that only transverse modes are excited, which strictly is valid
only for normal incidence. This assumption thus may break down for an angle of incidence of $\alpha$\,=\,11$^\circ$.
In the analysis of $R_p(\omega,\alpha,\varphi)$, the finite value of $\alpha$ and the mixing of longitudinal and transverse
modes are taken into account, see Eq.\ \ref{Rp}. The mixing of transverse and longitudinal character is particularly strong
if the corresponding eigenfrequencies are nearly degenerate.
The near degeneracy of $\omega_{\rm LO,7}$ and $\omega_{\rm 0,8}$ is apparent from Fig.\ \ref{MnWO4Epsilon},
which shows the two entries of the diagonal form of the real part Re$\{\hat{\varepsilon}_{ac}(\omega)\}$.
The two zero crossings of the diagonal components which correspond to $\omega_{\rm LO,7}$ and $\omega_{\rm 0,8}$
nearly coincide in frequency at roughly 760\,cm$^{-1}$ -- 770\,cm$^{-1}$.
This is the frequency range with the largest deviations between $R_{ac}(\omega,\chi)$ and the Lorentzian fit.
In this range, Re$\{\hat{\varepsilon}_{ac}(\omega)\}$ is close to zero in any direction within the $ac$ plane.
We propose that this causes the unusual line shape.

In $R_p(\omega,\alpha,\varphi)$ with $\varphi$\,=\,80$^\circ$, the highest $B_u$ mode predominantly shows longitudinal
character, giving rise to only a small peak close to $\omega_{\rm LO,8}$. The eigenfrequency of this predominantly
longitudinal mode does not coincide with $\omega_{\rm LO,7}$, thus the mixing of $B_u$ modes 7 and 8 does not play a
role for this geometry. Accordingly, $R_p(\omega,11^\circ,80^\circ)$ is well described by a model employing a
constant value of $\gamma_8$.

\begin{figure}[t!]
\includegraphics[width=0.75\columnwidth]{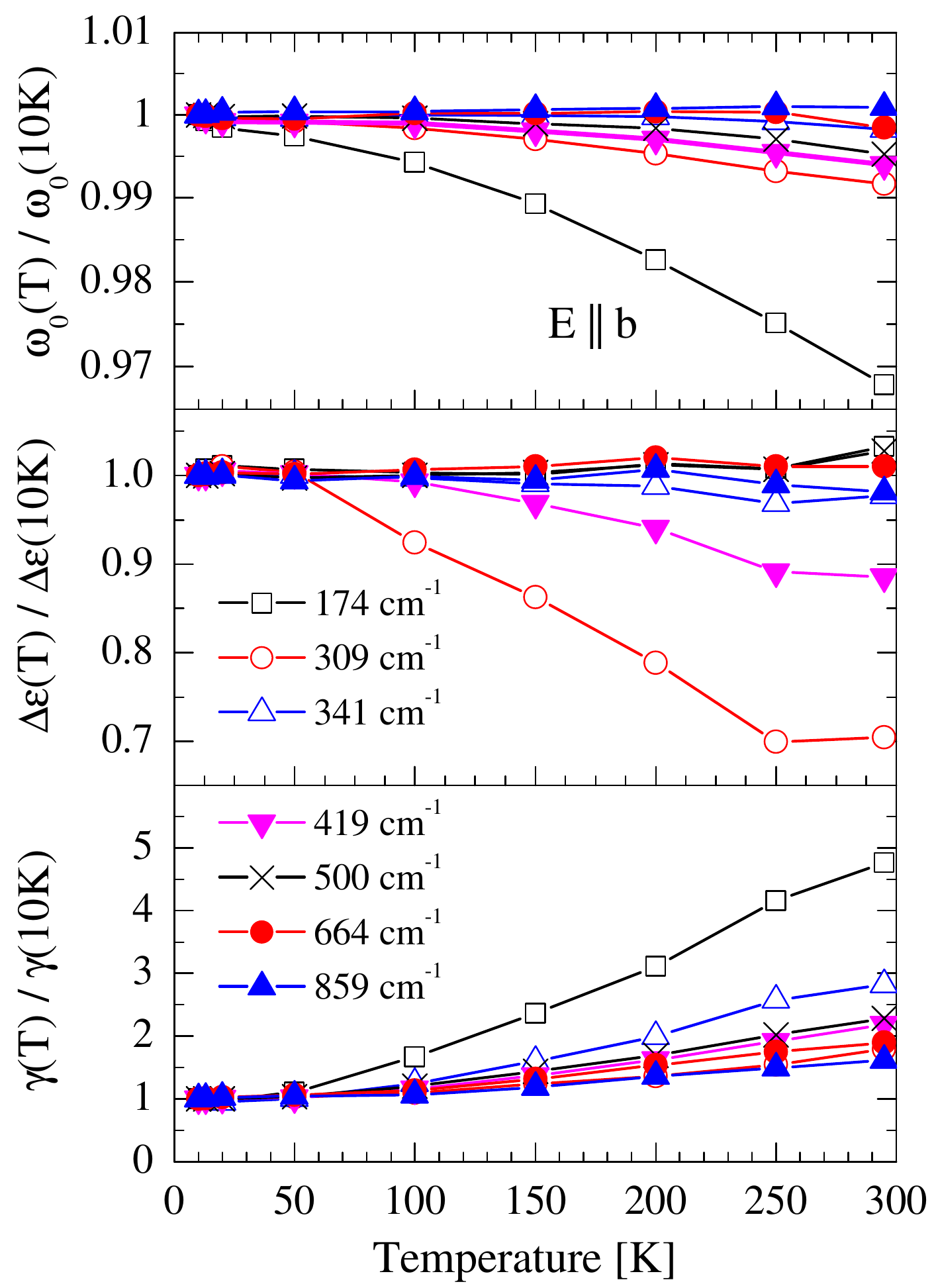}
\caption{(Color online) Temperature dependence of the parameters of the Drude-Lorentz fit for $E \, \| \, b$,
i.e., for the $A_u$ phonon modes:
eigenfrequency $\omega_{0,i}$, oscillator strength $\Delta\varepsilon_i$, and damping $\gamma_i$. }
\label{MnWO4RelParameterEbAll}
\end{figure}

\subsection{Temperature dependence}
\label{SecT}

\subsubsection{$A_u$ modes}

The Drude-Lorentz fit parameters of all $A_u$ phonon modes are plotted as a function of temperature
in Fig.\ \ref{MnWO4RelParameterEbAll}.
Above 20\,K, all parameters evolve smoothly with temperature, there is no evidence for any strong anomaly.
Six of the seven $A_u$ modes soften by only 1\,\% or less between 10\,K and 295\,K.\@
The lowest $A_u$ mode with $\omega_{0,1}$\,=\,174\,cm$^{-1}$ forms an exception. Between 10\,K and 295\,K,
it softens by 5.6\,cm$^{-1}$ or 3\,\%, showing the largest redshift for $E \, \| \, b$ both on a relative and
on an absolute scale. This pronounced shift can be clearly seen in the reflectivity data,
see Fig.\ \ref{MnWO4RefallT}.
In Raman data, the largest relative shift of 2.5\% between 5\,K and 300\,K is observed for the lowest $A_g$ mode.\cite{Iliev09}
Also in the case of the $B_u$ modes, the largest relative redshift is observed for the two lowest modes (see below).
The phonon softening observed above 20\,K can be attributed to the typical thermal expansion of the
lattice, reflecting anharmonicity.
The damping constants $\gamma_i$ of all $A_u$ modes behave as expected, showing a smooth increase with increasing
temperature. Again, the strongest change (on a relative scale) is observed for the lowest $A_u$ mode.
The oscillator strength $\Delta\varepsilon_i$, i.e., the contribution of a given mode $i$ to the real part of $\varepsilon_b$
at sub-phonon frequencies, is obtained via $\Delta\varepsilon_i$\,=\,$(\omega_{p,i}/\omega_{0,i})^2$.
On a relative scale, the $A_u$ mode at 309\,cm$^{-1}$ shows the most pronounced reduction of $\Delta\varepsilon_i$
with increasing temperature, but this mode also has the smallest absolute value of $\Delta\varepsilon_i$
(see Tab.\ \ref{MnWO4FitTable} and Fig.\ \ref{MnWO4RefallT}). According to Maczka \emph{et al.},\cite{Maczka11}
this mode corresponds to a WO$_2$ bending mode (twisting and scissoring,
see Tab.\ \ref{MnWO4FitTable}). Note that the observed phonon softening with increasing temperature leads to an \textit{enhanced}
oscillator strength $\Delta\varepsilon_i$. A \textit{decrease} of $\Delta\varepsilon_i$ with increasing temperature
reflects that the increase of $1/\omega_{0,i}^2$ is overcompensated by a reduction of the spectral weight
$\propto \omega_{p,i}^2$, i.e., of the effective ionic charge.

\begin{figure}[t!]
\includegraphics[width=0.9\columnwidth]{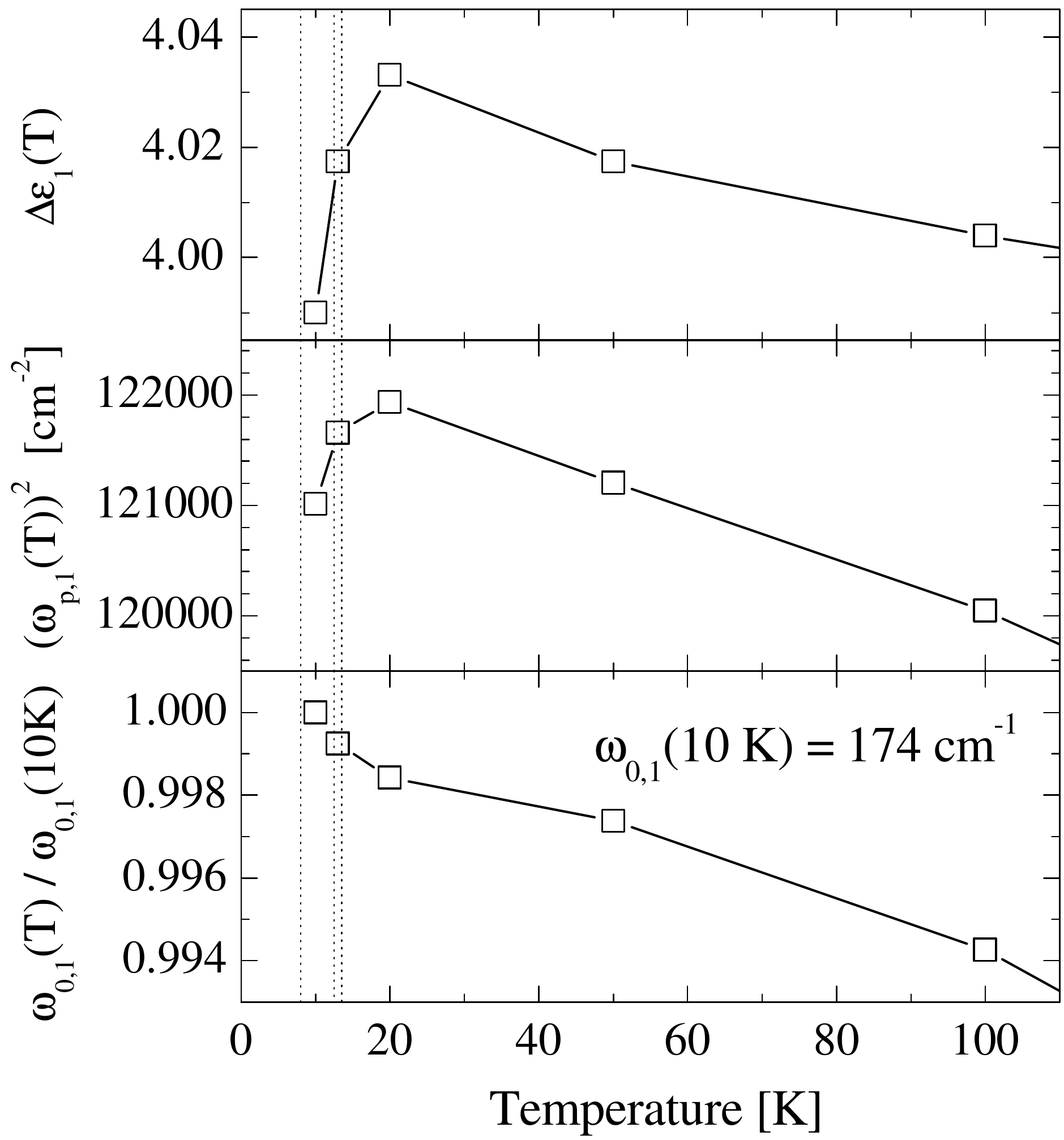}
\caption{(Color online) Eigenfrequency $\omega_{0,1}$
(bottom), spectral weight $\propto \omega_{p,1}^2$ (middle),
and oscillator strength $\Delta \varepsilon_1$ of the lowest $A_u$ phonon mode.
Dotted lines: phase transition temperatures. }
\label{MnWO4omega0Au1}
\end{figure}
\begin{figure}[t]
\includegraphics[width=0.85\columnwidth]{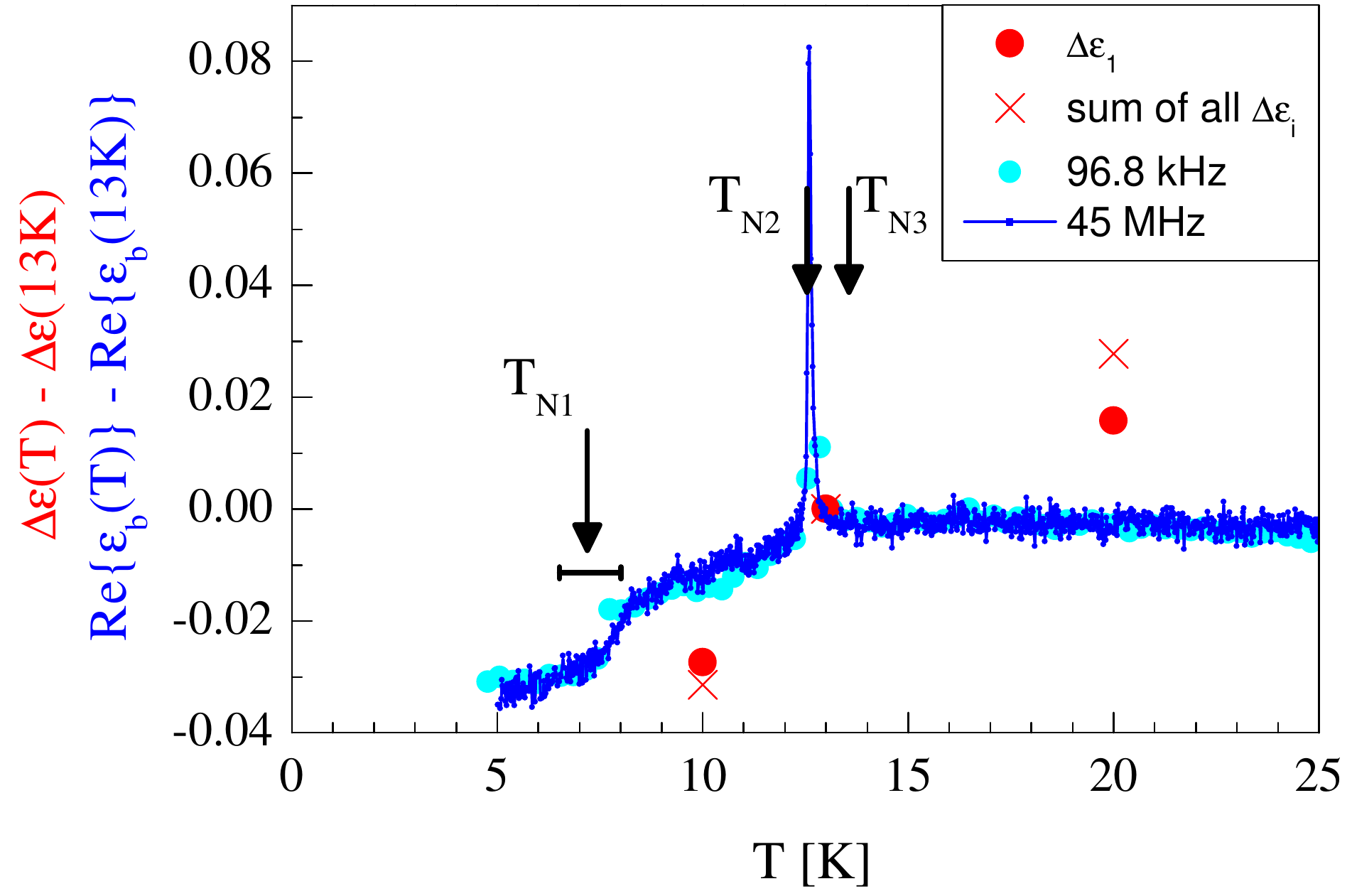}
\caption{(Color online) Blue and cyan blue symbols: Real part of the dielectric constant $\varepsilon_b(T)$ measured
by dielectric spectroscopy at 45\,MHz and 96.8\,kHz, respectively.
Red dots: Oscillator strength $\Delta\varepsilon_1$ of the lowest $A_u$ mode.
Red crosses: Sum of the oscillator strengths $\Delta\varepsilon = \Sigma_i \Delta\varepsilon_i$ of all $A_u$ modes.
All data sets are normalized to their value at 13\,K.\@ }
\label{MnWO4Au1}
\end{figure}

Below 20\,K, we observe only small changes of the phonon parameters of the $A_u$ modes.
We focus on the mode with $\omega_{0,1}$(10\,K)\,=\,174\,cm$^{-1}$, which exhibits the largest changes of $\omega_{0,i}$,
$\omega_{p,i}$, and $\Delta\varepsilon_i$ at low temperatures.
Figure \ref{MnWO4omega0Au1} shows the temperature dependence of
$\omega_{0,1}$, $\omega_{p,1}^2$, and of $\Delta\varepsilon_1$ on an enlarged scale.
The changes of $\omega_{0,1}$ between the different phases amount to only 0.1\,--\,0.2\,cm$^{-1}$ or 0.1\,\%,
which is hard to resolve experimentally. However, the values for $\omega_{0,1}$ found at 10\,K and 13\,K clearly deviate
from the approximately quadratic temperature dependence of $\omega_{0,1}$ observed above 20\,K.\@
In particular, the hardening of $\omega_{0,1}$ with decreasing temperature is the opposite of the
conventional phonon mode softening occurring in proper ferroelectrics.
At the same time, spin-phonon coupling is expected to cause a phonon shift proportional to the nearest-neighbor
spin-spin correlation function.\cite{Baltensperger68,Lockwood88,Sushkov05,Fennie06,Rudolf07}
For the spectral weight we find a reduction of
$\Delta\omega_{p,1}^2 = [\omega_{p,1}(13\,K)]^2-[\omega_{p,1}(10\,K)]^2 \approx 6\cdot 10^2$\,cm$^{-2}$,
see Fig.\ \ref{MnWO4omega0Au1}.
Note that this corresponds to only 0.5\,\% of $\omega_{p,1}^2$.
Together, the decrease of $\omega_{p,1}^2$ and the increase of $\omega_{0,1}^2$ between 13\,K and 10\,K
yield a reduction of the oscillator strength $\Delta\varepsilon_1$ of about 0.03
(see top panel of Fig.\ \ref{MnWO4omega0Au1} and red dots in Fig.\ \ref{MnWO4Au1}).
Note that the sum of the changes of all other modes including the change of $\varepsilon_\infty$
is about an order of magnitude smaller.

In the following, we estimate the oscillator strength of a possible electromagnon
by comparing the results of our phonon study with the temperature dependence of the real part Re$\{\varepsilon_b\}$
measured by dielectric spectroscopy at 45\,MHz and 96.8\,kHz, see Fig.\ \ref{MnWO4Au1}.
Consider a single infrared-active oscillator with oscillator strength $\Delta \varepsilon_i$, eigenfrequency $\omega_{0,i}$,
and damping constant $\gamma_i \ll \omega_i$. For $\omega \ll \omega_{0,i}$, this oscillator gives rise to a
constant contribution $\Delta \varepsilon_i$\,=\,$(\omega_{p,i}/\omega_{0,i})^2$ to Re$\{\varepsilon(\omega)\}$
(see, e.g., Eq.\ \ref{EPSb}).
For frequencies far below the phonon range, the contribution of phonons and higher-lying excitations to
Re$\{\varepsilon_b\}$ may thus be approximated by
Re$\{\varepsilon_b^{\rm high}\}$\,=\,$\varepsilon_{yy}^\infty + \Sigma_{i=1}^7 \Delta \varepsilon_i$,
where the sum is running over the seven $A_u$ modes. If Re$\{\varepsilon_b^{\rm high}\}$ is smaller than
the quasi-static value of Re$\{\varepsilon_b\}$, this implies a finite contribution of a further infrared-active
oscillator at intermediate frequencies.
We have chosen to measure Re$\{\varepsilon_b\}$ at 45\,MHz and 96.8\,kHz for two reasons.
First, these frequencies are far below the eigenfrequency $\omega_{\rm 0,em} \approx 2-3\,$cm$^{-1}$
of a possible electromagnon as observed in inelastic neutron scattering\cite{Braden,Ye11} and terahertz spectroscopy.\cite{PimenovPriv}
Secondly, they are high enough to neglect contributions from domain-wall dynamics.\cite{Niermann14}
Therefore, differences between Re$\{\varepsilon_b({\rm 45\,MHz})\}$ and the value of Re$\{\varepsilon_b^{\rm high}\}$
determined from the phonon parameters have to be attributed to the possible contribution of an electromagnon.

In Fig.\ \ref{MnWO4Au1}, we normalized both Re$\{\varepsilon_b({\rm 45\,MHz})\}$ and Re$\{\varepsilon_b^{\rm high}\}$
to their respective values at 13\,K in order to eliminate uncertainties of the absolute value.
At $T_{\rm N1}\approx 6.5 - 8.0$\,K the low-frequency data show a jump of Re$\{\varepsilon_b\}$ of about
0.01, in agreement with the results of Refs.\ [\onlinecite{Taniguchi06}] and [\onlinecite{Arkenbout06}].
The peak in the vicinity of $T_{\rm N2}$ reflects the washed-out divergence of Re$\{\varepsilon_b\}$
at the phase transition.
Between 13\,K and 10\,K, Re$\{\varepsilon_b\}$ {\it decreases} by about 0.01.
At first sight, this may seem to contradict the expected gain of oscillator strength of the electromagnon
upon entering the multiferroic phase.
The expected increase of Re$\{\varepsilon_b\}$ can be reconciled with the observed decrease of Re$\{\varepsilon_b\}$
by considering a decrease of the effective ionic charge causing a decrease of Re$\{\varepsilon_b^{\rm high}\}$.
As discussed above, we find that Re$\{\varepsilon_b^{\rm high}\}$ decreases by about 0.03 between 13\,K and 10\,K.\@
It is tempting to attribute the difference of 0.02 between the changes of Re$\{\varepsilon_b^{\rm high}\}$
and Re$\{\varepsilon_b\}$ to the oscillator strength $\Delta \varepsilon_{\rm em}$ of a possible electromagnon.
However, the subtle change of Re$\{\varepsilon_b^{\rm high}\}$ is difficult to quantify experimentally.
Hence we rather interpret this value as a rough estimate for the upper boundary of the electromagnon oscillator strength,
$\Delta \varepsilon_{\rm em} \lessapprox 0.02$.
Using the eigenfrequency $\omega_{0,{\rm em}}$\,$\approx$\,2-3\,cm$^{-1}$ of a possible electromagnon
as observed in inelastic neutron scattering and terahertz transmittance,\cite{PimenovPriv,Braden,Ye11}
we arrive at an estimate for the upper boundary of the electromagnon spectral weight
$\omega_{p,{\rm em}}^2$\,$\lessapprox$\,0.1-0.2\,cm$^{-2}$.
This value is tiny, it is about 6 orders of magnitude smaller than the oscillator strength $\omega_{p,1}^2$
of the lowest $A_u$ phonon mode, and still more than 3 orders of magnitude smaller than the
small reduction of $\omega_{p,1}^2$ observed between 13\,K and 10\,K.\@

From the point of view of symmetry, the oscillator strength of the electromagnon has to be finite
in the multiferroic phase. However, symmetry does not quantify the oscillator strength, it may be very small and
hard to detect experimentally.
In this context, it is interesting to note that the enhancement of the real part of the low-frequency dielectric
constant Re$\{\varepsilon_b\}$ directly above $T_{\rm N2}$ for frequencies below about 2\,GHz has been interpreted
as a signature of the critical slowing down of magnetoelectric fluctuations, i.e., softening of an overdamped
electromagnon.\cite{NiermannSoft}

We conclude that the spin-lattice interaction is not strong enough in MnWO$_4$ to change the phonon spectra
substantially at the magnetic phase transitions. Nevertheless we are able to resolve small changes of the $A_u$ mode
lowest in energy.
Note that also in the manganites $A$MnO$_3$ (with $A$\,=\,Gd$_{1-x}$Tb$_x$ and Eu$_{1-x}$Y$_x$)
it is the phonon mode lowest in energy that is most affected by the transition
to the multiferroic phase.\cite{Pimenov06,Takahashi08,Schmidt09,Schleck10,Lee09,Aguilar07,Takahashi09}
Contrary to the conventional phonon softening observed in proper ferroelectrics,
the hardening of phonons at the transition to the multiferroic phase may turn out to be a
characteristic property of multiferroics.\cite{Takahashi08,Schmidt09,Schleck10,Lee09}

\subsubsection{$B_u$ modes}

\begin{figure}[t]
\includegraphics[width=0.8\columnwidth,clip]{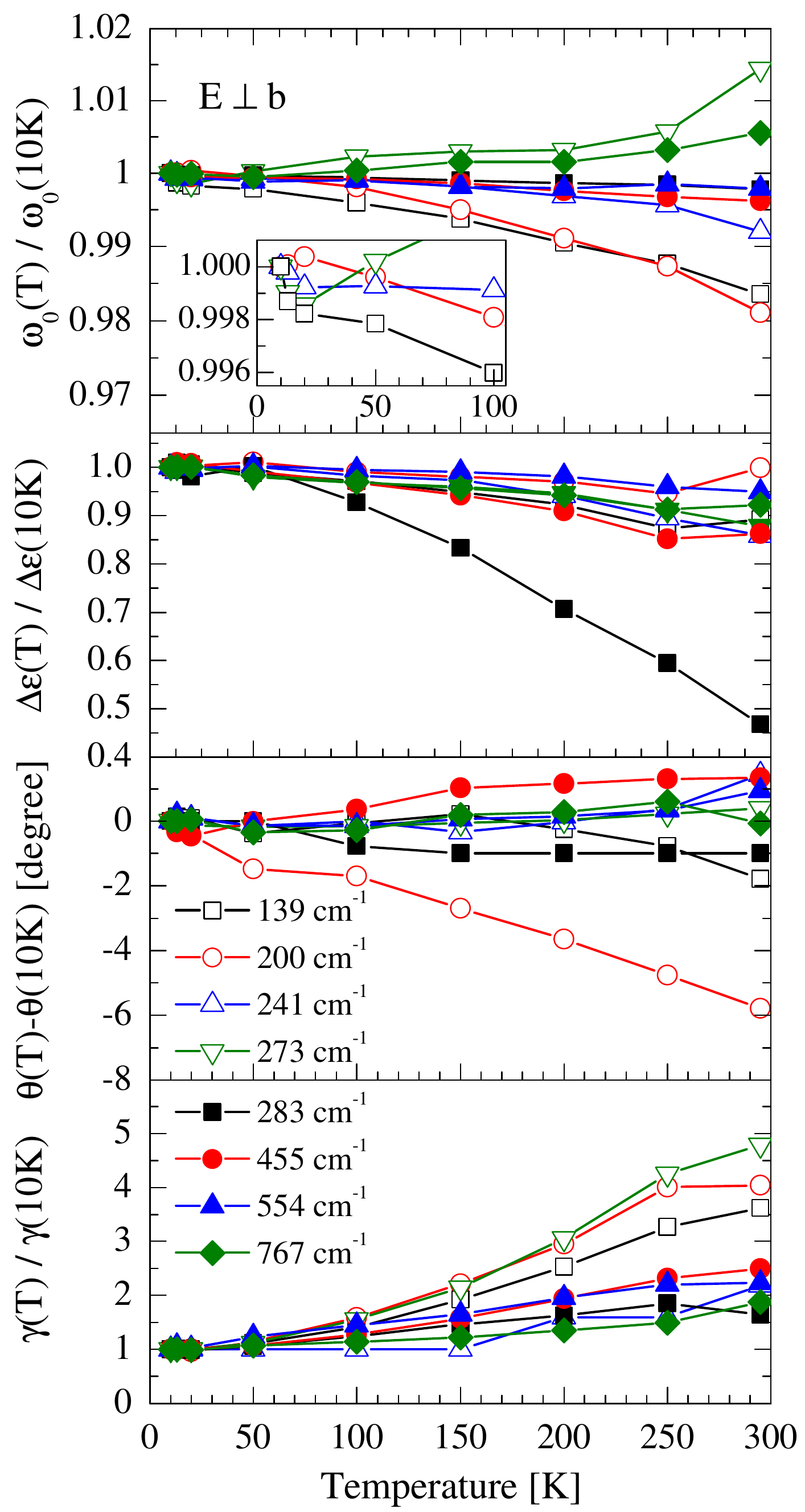}
\caption{(Color online) Temperature dependence of the parameters of the Drude-Lorentz fit for $E \! \perp \! b$,
i.e., for the $B_u$ phonon modes:
eigenfrequency $\omega_{0,i}$, oscillator strength $\Delta\varepsilon_i$, angle $\theta_i$, and damping $\gamma_i$.
Inset: Temperature dependence of $\omega_{0,i}$ for $i\,\in\,\{1-4\}$ on an enlarged scale. }
\label{MnWO4RelParameterEacAll}
\end{figure}

Figure \ref{MnWO4RelParameterEacAll} shows the temperature dependence of the parameters of the generalized Drude-Lorentz fit
for all $B_u$ phonon modes between 10\,K and 295\,K.\@
The overall picture is very similar to the case of the $A_u$ modes.
Six of the eight $B_u$ modes soften with increasing temperature, and the largest relative redshift between 10\,K and 295\,K
is observed for the two modes lowest in energy. The two modes with $\omega_{0,4}$\,=\,273\,cm$^{-1}$ and
$\omega_{0,8}$\,=\,767\,cm$^{-1}$ show an unexpected \emph{increase} of the eigenfrequency with increasing temperature.
For the highest $B_u$ mode with $\omega_{0,8}$\,=\,767\,cm$^{-1}$, we attribute this behavior to the fact
that the line shape of $R_{ac}(\omega)$ is not described very well by the generalized Drude-Lorentz model,
as discussed above. The deviations between fit and measured data for $R_{ac}(\omega, 30^\circ)$ at about 750\,cm$^{-1}$
[panel b) in Figs.\ \ref{MnWO4RefFit010K} and \ref{MnWO4RefFit295K}] show that $\omega_{0,8}$ cannot be determined
with the same precision as the eigenfrequencies of the other modes with a Lorentzian line shape.

The damping constants $\gamma_i$ of all $B_u$ modes increase with increasing temperature.
Similar to the $A_u$ modes, most $B_u$ modes show a modest increase of less than a factor of 2,
while the two modes lowest in energy show a more pronounced temperature dependence.
The temperature dependence of the rotation angles $\theta_i$ is only moderate, for seven modes the value of $\theta_i$
is stable within 2$^\circ$. A more detailed analysis of the rotation angles would require measurements for
a larger number of polarization directions $\chi$.
Finally, the temperature dependence of the oscillator strengths $\Delta\varepsilon_i$ is somewhat larger than
observed for the $A_u$ modes. Remarkably, the mode with $\omega_{0,5}$\,=\,283\,cm$^{-1}$ looses more than half
of its oscillator strength between 10\,K and 295\,K (see also Fig.\ \ref{MnWO4RefallT}).
As in the case of the $A_u$ modes, this strong relative change of $\Delta\varepsilon_i$ is observed for the
mode with the smallest value of $\Delta\varepsilon_i$.

Maczka \emph{et al.}\cite{Maczka11} observed anomalies in the eigenfrequencies and damping constants
of Mn$_{0.97}$Fe$_{0.03}$WO$_4$ and Mn$_{0.85}$Co$_{0.15}$WO$_4$ at about 50\,K and around 150 to 200\,K.\@
These results were obtained from absorption measurements on polycrystalline samples, which were grown
with a solvent, resulting in dark plates.\cite{Maczka11}
The dark color of these samples may be caused by the dopants Fe and Co, but it may also indicate that
not all Mn ions are in the divalent state, as it was found for flux-grown, undoped MnWO$_4$.\cite{Becker07}
In our ruby-red transparent MnWO$_4$ crystals, we do not find any evidence for anomalies above 20\,K,
neither for $B_u$ nor for $A_u$ phonon modes.

Below 20\,K, we find only subtle changes close to the experimental uncertainty.
For the $B_u$ modes, a precise determination of possible small changes of the oscillator strength $\Delta\varepsilon_i$
of oscillator $i$ is more difficult than for the $A_u$ modes because $\Delta\varepsilon_i$ is correlated to the
value of the orientation $\theta_i$.
The eigenfrequency $\omega_{0,1}$\,=\,139\,cm$^{-1}$ of the lowest $B_u$ mode hardens by about 0.05\% between 20\,K and 13\,K
and by about 0.13\% across $T_{\rm N2}$ between 13\,K and 10\,K (see the inset of Fig.\ \ref{MnWO4RelParameterEacAll}).
These values are very similar to the results obtained for the lowest $A_u$ mode.

\section{Conclusion}

We report on the first analysis of the full dielectric tensor in the frequency range of the phonons
for any mono\-clinic tungstate $A$WO$_4$ with divalent $A$ metal ions.
In MnWO$_4$, we unambiguously identified all infrared-active phonon modes (7 $A_u$ modes and 8 $B_u$ modes)
and determined their temperature dependence.
In particular the strongest $B_u$ modes were overlooked in previous studies.
A full polarization analysis is an essential prerequisite for the identification of the correct number of modes
and of their parameters, in particular in the case of overlapping modes.
For phonon modes with $B_u$ symmetry, the combined analysis of $R_p(\omega)$ -- measured with
the (010) plane as plane of incidence -- and $R_{ac}(\omega,\chi)$ is best suited for a reliable determination of
$\hat{\varepsilon}_{ac}$ and of the phonon parameters.
In the data on $R_{ac}(\omega,\chi)$ of MnWO$_4$, we found deviations of the expected Lorentzian line shape
around 750\,cm$^{-1}$ in a frequency range where both Re$\{\varepsilon_{xx}\}$ and Re$\{\varepsilon_{zz}\}$
are close to zero.
We derived a Kramers-Kronig-consistent oscillator model which is able to describe
asymmetric line shapes in compounds with monoclinic symmetry.
In particular, our model also describes modes showing a gradual rise
of Im$\left\{ \varepsilon(\omega) \right\}$ at the low-frequency side and a steep drop at high energies.
We propose that the asymmetric line shape observed in $R_{ac}(\omega,\chi)$ for near-normal incidence
is caused by the mixing of longitudinal and transverse modes in a frequency range where
Re$\{\varepsilon_{ac}\} \approx 0$ for all directions within the $ac$ plane.

Based on a generalized Drude-Lorentz model, we determined the temperature dependence of the phonon parameters,
including the orientation of the $B_u$ modes within the $ac$ plane.
The phonons show only subtle changes at the magnetic phase transitions between the AF2, AF3, and
the paramagnetic phase.
In particular, the eigenfrequency of the lowest $A_u$ mode \textit{increases} by roughly 0.1\%
upon cooling from 13\,K to 10\,K, while the spectral weight decreases by
$(\Delta\omega_{p,1}/\omega_{p,1})^2$\,$\approx$\,0.5\%.
A comparison to the data for the quasi-static dielectric constant Re$\{\varepsilon_b\}$ yields an upper boundary
for the spectral weight of a possible electromagnon of less than 0.1\% of the small change $\Delta\omega_{p,1}^2$
of the spectral weight of the lowest $A_u$ mode.
Phonon hardening upon entering the multiferroic phase may turn out to be a
characteristic property of magnetoelectric multiferroics of spin-spiral type.
In contrast to previous reports on MnWO$_4$ or slightly doped Mn$_{1-x}$A$_x$WO$_4$,
we do not find any anomalies above 20\,K.\@
We conclude that spin-lattice coupling in MnWO$_4$ is only small.

The quantitative understanding of the phonon modes obtained here provides an excellent starting point for the
analysis of the optical data of nanocrystals\cite{Maczka11b,Maczka12,Tong10}
as well as for an attempt to comprehend the anomalies reported for doped polycrystalline samples.\cite{Maczka11,Ptak12}

It is a pleasure to acknowledge fruitful discussions with P.H.M. van Loosdrecht.
This work was supported by the DFG via SFB 608.

%-------------------------------------------------------------------------------

\end{document}